\documentclass[showpacs,nofootinbib,preprintnumbers,twocolumn,amsmath,amssymb]{revtex4}

\usepackage{datetime}
\usepackage{color}

\usepackage{graphicx}
\usepackage{slashed}
\usepackage{bbm}
\usepackage{footmisc}

\renewcommand{\thefootnote}{\fnsymbol{footnote}}
\newcommand{\rig}{\rightarrow}

\renewcommand{\d}{{\mathrm{d}}}
\newcommand{\be}{\begin{eqnarray*}}
\newcommand{\ee}{\end{eqnarray*}}
\newcommand{\gl}[1]{(\ref{#1})}

\newcommand{\bee}{\begin{eqnarray}}
\newcommand{\eee}{\end{eqnarray}}
\newcommand{\beeq}{\begin{equation}}
\newcommand{\eeeq}{\end{equation}}
\newcommand{\gev}{~{\rm{GeV}}}
\newcommand{\tev}{~{\rm{TeV}}}

\newcommand{\ep}{\varepsilon}
\renewcommand{\vec}{\bf}
\newcommand{\emt}{$\times 10^{-3}$}
\newcommand{\emfo}{$\times 10^{-4}$}
\newcommand{\emfi}{$\times 10^{-5}$}

\begin{document}

\title{Measuring spin and ${\cal{CP}}$ from semihadronic $ZZ$ decays using jet substructure}

\begin{abstract}
We apply novel jet techniques to investigate the spin and ${\cal{CP}}$
quantum numbers of a heavy resonance $X$, singly produced in 
$pp\rig X \rig ZZ \rig \ell^+\ell^- jj$ at the LHC. We take into account all 
dominant background processes to show that this channel, which has been considered unobservable until now,
can qualify under realistic conditions to supplement measurements of the purely leptonic decay channels
$X\rig ZZ \rig 4\ell$.
We perform a detailed investigation of spin- and ${\cal{CP}}$-sensitive angular observables 
on the fully-simulated final state for various spin and ${\cal{CP}}$ quantum 
numbers of the state $X$, tracing how potential sensitivity communicates through 
all the steps of a subjet analysis. This allows us to elaborate on the prospects and
limitations of performing such measurements with the semihadronic final state. 
We find our analysis particularly sensitive to a ${\cal{CP}}$-even or 
${\cal{CP}}$-odd scalar resonance, while, for tensorial and vectorial resonances, discriminative
features are diminished in the boosted kinematical regime.
\end{abstract}

\author{Christoph Englert}
\email{c.englert@thphys.uni-heidelberg.de}
\affiliation{Institute for Theoretical Physics, Heidelberg University, 69120 Heidelberg, Germany}
\author{Christoph Hackstein}
\email{christoph.hackstein@kit.edu}
\affiliation{Institute for Experimental Nuclear Physics, Karlsruhe Institute of Technology, 76131 Karlsruhe, Germany}
\affiliation{Institute for Theoretical Physics, Karlsruhe Institute of Technology, 76131 Karlsruhe, Germany}
\author{Michael Spannowsky}
\email{mspannow@uoregon.edu}
\affiliation{Institute of Theoretical Science, University of Oregon, Eugene, Oregon 97403-5203, USA}

\preprint{IEKP--KA/2010--21 \;\;\; KA--TP--28--2010}

\maketitle

\section{Introduction}
\label{sec:intro}
An experimental hint pointing towards the source of electroweak symmetry breaking (EWSB) remains missing.
From a theoretical perspective, unitarity constraints do force us to expect new physics manifesting 
at energy scales $\lesssim 1.2 \tev$. For most of the realistic new physics models, this effectively means the 
observation of at least a single new resonant state in the weak boson scattering (sub)amplitudes, which serves
to cure the bad high-energy behavior of massive longitudinal gauge boson scattering. An equal constraint does also
follow from diboson production from massive quarks, $q\bar q\rig W^+W^-$, which relates EWSB to the dynamics
of fermion mass generation \cite{cornwall}. Revealing the unitarizing resonance's properties, such 
as mass, spin and ${\cal{CP}}$ quantum numbers, is a primary goal of the CERN Large Hadron Collider (LHC)
\cite{Aad:2008zzm,Ball:2007zza}. Completing this task will provide indispensable information 
elucidating the realization of EWSB and will therefore help to pin down the correct mechanism of 
spontaneous symmetry breaking.

At the LHC, the experimentally favorite channel to determine ${\cal{CP}}$ and spin of a new massive 
particle $X$ (\mbox{$m_X\gtrsim 300 \gev$}), coupling to weakly charged gauge bosons, is its decay to four charged leptons 
via \mbox{$X\rig ZZ$}~\mbox{\cite{Buszello:2002uu,Buszello:2004uu,Gao:2010qx,DeRujula:2010ys,Choi:2002jk}}. 
These ``golden channels'' \cite{goldplatedmode} are characterized by extraordinarily clean signatures, making them 
experimentally well-observable at, however, small rates due to the small leptonic $Z$ branching ratios. Until now, 
$X\rig ZZ$ for the semihadronic decay channels $ZZ\rig \ell^+ \ell^-jj$ has been considered experimentally disfavored.
In part, this is due to the overwhelmingly large backgrounds from underlying event and quantum chromodynamics 
(QCD), which exceed the expected signal by a few orders of magnitude.
An analogous statement has been held true for associated standard model Higgs production, until, only 
very recently, new subjet techniques have proven capable of discriminating signal from background for highly boosted Higgs kinematics 
\cite{Butterworth:2008iy,Plehn:2009rk,Soper:2010xk}. Subsequently, these phase space regions have received lots of attention 
in phenomenological analysis, demonstrating the capacity of subjet-related searches at the LHC in various channels and models \cite{subjets}. $H \rig WW$ was one of the first channels subjet techniques were ever applied to \cite{Seymour:1993mx}.
The purpose of these analysis was predominantly to isolate a resonance peak from an invariant mass distribution. In this paper we 
use a combination of different subjet methods for a shape-analysis of spin- and ${\cal{CP}}$-sensitive observables in a semihadronic
final state. ${\cal{CP}}$ properties of the standard model (SM)-like associated Higgs production have been investigated recently in 
Ref.~\cite{Christensen:2010pf} using $b$-tagging in $HZ\rig \ell^+ \ell^-b \bar b$.
 
Subjet techniques have also proven successful in $pp\rig X \rig ZZ \rig \ell^+\ell^-jj$ for standard model-like Higgs production 
($X=H$ with $m_H\gtrsim 350\gev$) in Ref.~\cite{Hackstein:2010wk}, yielding a $5\sigma$ discovery reach for an integrated luminosity 
of $10~{\rm{fb}}^{-1}$, when running the LHC at a center-of-mass energy of $\sqrt{s}=14\tev$. Equally important, the semihadronic decay 
channel exhibits a comparable statistical significance as the purely leptonic channels $pp \rig X \rig ZZ \rig \ell\ell\ell'\ell'$.
This can be of extreme importance if the LHC is not going to reach its  center-of-mass design-energy.
Hence, there is sufficient potential to revise semihadronic decays, 
not only to determine the resonance mass, but also its spin- and ${\cal{CP}}$ properties. 

The main goal of this paper is to 
investigate the attainable extent of sensitivity to the spin and 
${\cal{CP}}$ quantum numbers of a resonance $X$ in the channel $pp \rig  X \rig ZZ \rig\ell^+\ell^- jj $,
for the selection cuts, which allow to discriminate the signal from the background.
To arrive at a reliable assessment, we take into account realistic simulations of both the signal and the dominating background 
processes. 
We fix the mass and the production
modes of $X$, as well as its production cross section to be similar to the SM Higgs boson 
expectation\footnote{We normalize the cross section to SM Higgs production at
the parton level.}. 
On the one hand, this approach
can be motivated by again referring to unitarity
constraints: Curing the growth of both the $VV\rig VV$ and $q\bar q \rig WW$ scattering amplitudes by a
\emph{singly dominating} additional resonance fixes the overall cross section
to be of the order of the SM
(see e.g. \cite{Duhrssen:2004cv,Birkedal:2004au} for non trivial examples). 
On the other hand, we would like to focus on an experimental situation, which favors
the SM expectation, but leaving ${\cal{CP}}$ and spin properties as an
open question. For this reason, we also do not include additional dependencies of the
cross section on the width of $X$. 
The width is, in principle, an additional, highly model-dependent parameter, which can be vastly different from the
SM Higgs boson width (e.g. in models with EWSB by strong interactions~\cite{Chanowitz:1985hj,silh} 
or in so-called hidden-valley models \cite{Strassler:2006im}).
Instead, we straightforwardly adopt the SM Higgs boson width, which then turns the resonance
considered in this paper into a ``Higgs look-alike'', to borrow the language of Ref.~\cite{DeRujula:2010ys}.

%
%--------------------------------------------------------------------------------------------------------------------------------------------------------
% Figure CP angles
%--------------------------------------------------------------------------------------------------------------------------------------------------------
\begin{figure}[b!]
\includegraphics[width=0.48\textwidth]{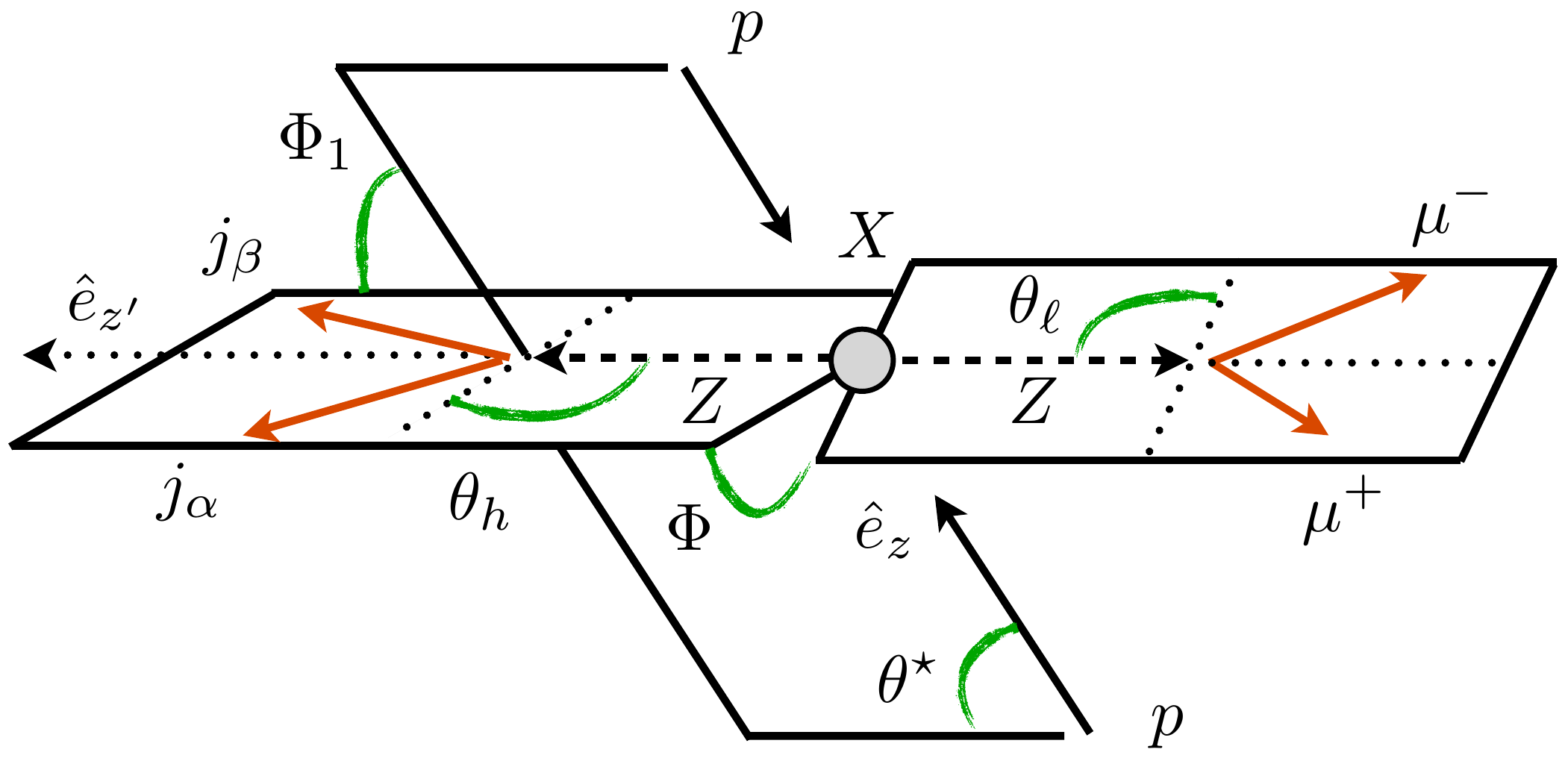}
\caption{\label{fig:cpangles} Spin- and ${\cal{CP}}$-sensitive angles of Ref.~\cite{Dell'Aquila:1985ve} 
in $pp\rig X \rig ZZ \rig \mu^+\mu^- jj$. Details on the angles' definition and on the assignment 
of $j_\alpha$ and $j_\beta$ are given in the text. An angle analogous to $\Phi_1$ can be defined with respect
to the leptonic decay plane. We refer to this angle as $\tilde\Phi$.}
\end{figure}

We organize this paper as follows:
In Sec.~\ref{sec:details}, we outline the necessary technical details of our analysis.
We review the effective interactions, from which we compute the production 
and the decay of the resonance $X$ with quantum numbers $J^{CP}=0^\pm,1^\pm,2^+$.
We also comment on the signal and background
event generation and the chosen selection criteria, and we introduce the ${\cal{CP}}$ 
and spin-sensitive observables and their generalization to semihadronic final states.
We discuss our numerical results in Sec.~\ref{sec:results}; 
Sec.~\ref{sec:conc} closes with a summary and gives our \mbox{conclusions}.

\section{Details of the analysis}
\label{sec:details}
\subsection{Spin- and ${\cal{CP}}$-sensitive observables}
The spin and ${\cal{CP}}$ properties are examined through correlations in the angular distributions of
the decay products. 
A commonly used (sub)set of angles is given by the definitions of 
Cabibbo and Maksymowicz of Ref.~\cite{Cabibbo:1965zz}, which originate from similar 
studies of the kaon system (see, e.g., Refs.~\cite{DeRujula:2010ys,Buszello:2002uu,Bredenstein:2006rh,Cao:2009ah} 
for their application to the $X\rig ZZ$). In this paper we focus on the angles of Ref.~\cite{Dell'Aquila:1985ve}
as sensitive observables, which also have been employed in the recent $X\rig 4l$ investigation in Ref.~\cite{Gao:2010qx}.
We quickly recall their definition with the help of Fig.~\ref{fig:cpangles}: Let ${\vec{p}}_\alpha,~{\vec{p}}_\beta$, and  ${\vec{p}}_\pm$ be the three-momenta
of the (sub)jets $j_\alpha$ and $j_\beta$ and the leptons in the laboratory frame, respectively. 
From these momenta, we compute the three-momenta of the hadronically and leptonically decaying $Z$ bosons
\begin{subequations}
\label{eq:angledefinition}
\beeq
{\vec{p}}_{Z_h} = {\vec{p}}_{\alpha} + {\vec{p}}_{\beta}\,, \qquad
{\vec{p}}_{Z_\ell} = {\vec{p}}_{+} + {\vec{p}}_{-}\,,
\eeeq
as well as the lab-frame $X$ three-momentum
\beeq
{\vec{p}}_{X}={\vec{p}}_{\alpha} + {\vec{p}}_{\beta}+ {\vec{p}}_{+} + {\vec{p}}_{-}\,.
\eeeq
In addition, we denote the normalized unit vector along the beam axis measured in the $X$ rest frame by $\hat e_z$, and the unit
vector along the $ZZ$ decay axis in the $X$ rest frame by $\hat e_{z'}$. 
The angles of Fig.~\ref{fig:cpangles} are then defined as follows
\begin{align}
\label{eq:costhetahel}
\cos\theta_h =&   { {\vec{p}}_\alpha \cdot {\vec{p}}_{X}  \over  \sqrt{  {\vec{p}}^2_\alpha\, {\vec{p}}^2_{X}    }} \bigg|_{Z_h} ,
&\hspace{-0.2cm}
\cos\theta_\ell =&   { {\vec{p}}_- \cdot {\vec{p}}_{X}  \over \sqrt{  {\vec{p}}^2_-\, {\vec{p}}^2_{X}    }} \bigg|_{Z_\ell} ,\\
\label{eq:costhetastar}
\cos\theta^\star =&   { {\vec{p}}_{Z_\ell} \cdot {\hat e_{z'}}  \over \sqrt{  {\vec{p}}^2_{Z_\ell} }} \bigg|_{X}, 
&\hspace{-0.2cm}
\cos\tilde\Phi =&  { (\hat e_z\times \hat e_{z'})\cdot ({\vec{p}}_- \times {\vec{p}}_+) \over 
 \sqrt{( {\vec{p}}_- \times {\vec{p}}_+ )^2} } \bigg|_X ,
\end{align}
\vspace{-0.5cm}
\beeq
\label{eq:decayplane}
\cos\Phi = { ( {\vec{p}}_\alpha \times {\vec{p}}_\beta ) \cdot ( {\vec{p}}_- \times {\vec{p}}_{+} ) \over
\sqrt{( {\vec{p}}_\alpha \times {\vec{p}}_\beta )^2\,  ( {\vec{p}}_- \times {\vec{p}}_{+} )^2 }}  \bigg|_X\,,
\eeeq
\end{subequations}
where the subscripts indicate the reference system, in which the angles are evaluated.
More precisely, the helicity 
angles $\theta_h$ and $\theta_\ell$ are defined in their mother-$Z$'s rest frame, and all other angles 
are defined in the rest frame of the particle $X$, where ${\vec{p}}_{Z_\ell}=-{\vec{p}}_{Z_h}$.
It is also worth noting that the helicity angles correspond to the so-called Collins-Soper angle 
of Ref.~\cite{Collins:1977iv}, evaluated for the respective $Z$ boson.

There is a small drawback when carrying over the definitions of Eq.~\gl{eq:angledefinition} from the purely leptonic decay channels
to the considered semihadronic final state: When dealing with $X\rig \ell^+\ell^- \ell'^+ \ell'^-$, it is always
possible to unambiguously assign a preferential direction for the lepton pairs by tagging their 
charge\footnote{Charge tagging can be considered ideal for our purposes, given the additional 
uncertainties from parton showering; see Sec.~\ref{sec:results}. For the region in pseudorapidity
and transverse momentum that we consider, the mistagging probability of, e.g., muons is typically
at the level of 0.5\%, even for early data-taking scenarios~\cite{Bayatian:2006zz}.}. 
This allows us to fix a convention for the helicity
angles, as well as for the relative orientation of the decay planes via a specific order of the three-momenta
when defining the normal vectors in Eq.~\gl{eq:angledefinition}. Considering semihadronic
$X$ decays, we are stuck with a twofold ambiguity, which affects the angular distributions. Even worse, 
$p_T$-ordered hard subjets,
dug out from the ``fat jet'' during the 
subjet analysis (for details see Sec.~\ref{sec:discr}) can bias the distributions. Hence, we need to impose 
an ordering scheme which avoids these shortcomings. 
An efficiently working choice on the inclusive parton level is provided by the imposing rapidity-ordering
\bee
\label{eq:raporder}
y(j_{\alpha}) < y(j_{\beta})\,,
\eee
which is reminiscent of the ${\cal{CP}}$-sensitive $\Phi_{jj}$ observable in vector boson 
fusion \cite{Hankele:2006ma}. This choice, however, does not remove all ambiguities. The orientation
of the decay planes [Eq.~\gl{eq:decayplane}] is not fixed by ordering the jets according to Eq.~\gl{eq:raporder}. 
The unresolved ambiguity results in averaging $\cos\Phi$ and $\cos(\pi-\Phi)$ over the event 
sample, leaving a decreased sensitivity in the angle $\Phi\in [0,\pi]$. We discuss this in 
more detail in Sec.~\ref{sec:results}.

\subsection{Simulation of signal and background events}
We generate signal events \mbox{$pp\rig X \rig ZZ \rig \ell^+ \ell^-jj$} with \textsc{MadGraph/MadEvent} \cite{Alwall:2007st}, 
which we have slightly modified to fit the purpose of this work. In particular these modifications include supplementing 
additional {\textsc{Helas}} \cite{Murayama:1992gi} routines and modifications of the \textsc{MadGraph}-generated 
code to include vertex structures and subprocesses that are investigated in this paper.
We have validated our implementation against existing spin correlation results of Refs.~\cite{Djouadi:2005gi,Gao:2010qx}. 
We choose the partonic production modes to be 
dependent on the quantum numbers of the particle $X$:
\begin{subequations}
\label{eq:subprocs}
\begin{align}
X=0^\pm: & \quad g g \rig X \rig ZZ \rig \ell^+ \ell^- jj\,, \\
X=1^\pm: & \quad q\bar q \rig X \rig ZZ \rig \ell^+ \ell^- jj\,, \\
X=2^+:& \quad gg \rig X \rig ZZ \rig \ell^+ \ell^- jj\,,
\end{align}
\end{subequations}
where $g$ denotes the gluon and $q,j=(u,d,s,c)$ represents the light constituent 
quarks of the proton.

The bottom quark contributions are negligibly small. While, in the light of the 
effective theory language of Ref.~\cite{Gao:2010qx}, this specific choice can 
be considered as a general assumption of our analysis, the partonic subprocesses
of Eq.~\gl{eq:subprocs} reflect the dominant production modes at the LHC. In particular,
the production of an uncolored vector particle $1^-$ from two gluons via fermion loops is forbidden
by Furry's theorem \cite{furrythrm}, while a direct $ggZ'$ coupling is ruled out by Yang's theorem \cite{Yang:1950rg}.

The effective operators that we include for the production and the decay of $X$ 
do not exhaust all possibilities either (see again Ref.~\cite{Gao:2010qx} 
for the complete set of allowed operators). Yet, we adopt a general enough set of operators
to adequately highlight the features of objects $X$ with different spins and ${\cal{CP}}$ quantum numbers in
our comparative investigation in Sec.~\ref{sec:results}.
The effective vertex function, from which we derive the
effective couplings of $X$ to the SM $Z$ bosons, that appear in the calculation of
the matrix elements in Eq.~\gl{eq:subprocs}, 
reads for the scalar case suppressing the color indices \cite{Hagiwara:1986vm},
\begin{subequations}	
\label{eq:decdens}
\beeq
{\cal{L}}^{ZZX}_{\mu\nu}=c^s_1\, g_{\mu\nu} + {c_2^s \over m_Z^2}  \epsilon_{\mu\nu\rho\delta}\, p_1^\rho \, p_2^\delta\,.
\eeeq
For a vectorial $X$, the vertex function follows from the generalized Landau-Yang theorem \cite{Keung:2008ve}
\begin{multline}
\label{eq:vecdec}
{\cal{L}}^{ZZX}_{\mu\nu\rho}=  c^v_1\left( g_{\mu \rho}\, p_{1,\nu} + g_{\nu \rho}\, 
       p_{2,\mu} \right) \\  -c^v_2\, \epsilon_{\mu\nu\rho\delta} \left( p_{1}^{\delta} - p_{2}^{\delta} \right ),
\end{multline}       
while for tensorial $X$, we include the vertex function \cite{spin2}
\begin{multline}
\label{eq:tensdec}
{\cal{L}}^{ZZX}_{\mu\nu\rho\delta}= c_1^t  \big( 
p_{1,\nu}\, p_{2,\rho}\, g_{\mu\delta} + p_{1,\rho}\, p_{2,\mu} \,g _{\nu\delta}
\\  + p_{1,\rho}\, p_{2,\delta}\, g_{\mu\nu}  -{1\over 2} m_X^2\, g_{\mu\rho}\, g_{\mu \delta} \big)
\end{multline}
\end{subequations}
to our comparison.

From Eqs.~\gl{eq:decdens}, we can determine the (off-shell) decays 
$X,X_\rho,X_{\rho\delta} \rig Z_\mu(p_1) Z_\nu(p_2)$ by contracting with
the final state $Z$ bosons' effective polarization vectors $\ep^\ast_\mu(p_1)$ and $\ep^\ast_\nu(p_2)$,
which encode the Breit-Wigner propagator and the respective $Z$ decay vertex.
%
%--------------------------------------------------------------------------------------------------------------------------------------------------------
% Efficiency table
%--------------------------------------------------------------------------------------------------------------------------------------------------------
\begin{table*}[t]
\setcounter{table}{1}
\begin{tabular}{c || c  c | c  c |  c  c | c  c | c  c || c c | c c | c c}
\hline
  &	\multicolumn{2}{c |}{$0^+$}  & \multicolumn{2}{c |}{$0^-$} & \multicolumn{2}{c|}{$1^-$} & \multicolumn{2}{c|}{$1^+$} &
  \multicolumn{2}{c||}{ $2^+$} & \multicolumn{2}{c|}{$Z+\rm{jets}$} & \multicolumn{2}{c|}{$ZZ$} &
  \multicolumn{2}{c}{ $t\bar t$}\\	
\hline
\hline
	& P & H & P & H& P & H& P & H& P & H & P & H& P & H& P & H \\
\hline
Raw	& \multicolumn{2}{c|}{1.00} & \multicolumn{2}{c|}{1.00} & \multicolumn{2}{c|}{1.00} & \multicolumn{2}{c|}{1.00} & \multicolumn{2}{c||}{1.00}  & \multicolumn{2}{c|}{1.00} & \multicolumn{2}{c|}{1.00} & \multicolumn{2}{c}{1.00}	\\
Cuts	                     	& 0.41	& 0.53 	& 0.35 	& 0.47 	& 0.28 	& 0.40 	& 0.29	& 0.42 	& 0.31 	& 0.40 	& 0.15 & 0.17 & 0.24 & 0.36 & 0.02 & 0.01\\
Hadr.$Z$ 			& 0.22 	& 0.29 	& 0.16 	& 0.22 	& 0.16 	& 0.22 	& 0.16 	& 0.23  	& 0.15 	& 0.19 	&  4.2\emt & 6.5\emt & 0.02 & 0.03 & 1.2\emt & 0.8\emt\\
$m_X$     			& 0.17  	& 0.22 	& 0.12 	& 0.16 	& 0.12 	& 0.17 	& 0.13 	& 0.18 	& 0.11 	& 0.15 	& 1.6 \emt & 2.2\emt & 4.7\emt &  7.0\emt & 2.3\emfo & 1.6\emfo \\
$\Delta R_{ZZ}$ 	& 0.15 	& 0.20 	& 0.11 	& 0.14  	& 0.10 	&0.14  	& 0.10 	& 0.14 	& 0.10 	& 0.13 	& 1.3 \emt & 1.9 \emt & 3.7\emt & 5.7 \emt & 2.0\emfo & 1.3\emfo \\ 	
Tr+Pr   			& 0.10 	& 0.13 	& 0.07 	& 0.09 	& 0.07  	&0.10 	& 0.07 	& 0.10 	& 0.06 	& 0.08 	& 4.7 \emfo & 5.7\emfo & 1.9\emt & 2.9\emt & 7.8\emfi & 4.2\emfi \\
\hline
\hline
$\sigma$ [fb] & 	 \multicolumn{2}{c|}{11.7} & \multicolumn{2}{c|}{8.3} & \multicolumn{2}{c|}{8.7} & \multicolumn{2}{c|}{9.1} & \multicolumn{2}{c||}{7.5}  & \multicolumn{6}{c|}{29.7} \\
\hline
\end{tabular}
\caption{\label{tab:cutscenarios} Cut flow comparison of the {\sc{MadEvent}} signal event when processed either with
{\sc Pythia} 6.4 (referred to as P) or {\sc Herwig++}  (denoted by H)
for the $X$ states of Table~\ref{tab:scenarios}.
Starting from the showered sample on calorimeter level (Raw), we apply the selection cuts
(Cuts), the hadronic $Z$ reconstruction requirements, the $X$ mass reconstruction ($m_X$),
the $S/B$-improving requirement on $\Delta R_{ZZ}$, and trimming and pruning (Tr+Pr).
The selection criteria are described in detail in Sec.~\ref{sec:discr}. The lower row gives the total cross section
determined from the {\sc{Herwig++}} efficiencies after all cuts have been applied, cf. \cite{Hackstein:2010wk}. The different
acceptance levels follow from the different angular correlations, see Sec.~\ref{sec:discr}. \vspace{-0.2cm}}
\end{table*}
We include the spin and $\cal{CP}$ dependence of the $X$ production from quarks 
via the effective Lagrangian in the vectorial scenario \cite{Murayama:1992gi} 
\begin{subequations}
\label{eq:proddens}
\beeq
{\cal{L}}^{q\bar qX}=  \bar\Psi_q \gamma^\mu\left( g_L^v\, {\mathbb{P}}_L + g_R^v \, {\mathbb{P}}_R \right) \Psi_q\,
 X_\mu\,,
\eeeq
where
\beeq
{\mathbb{P}}_{L,R}={1\over 2} \left( {{\mathbbm{1}}} \mp \gamma_5 \right)\,,
\eeeq
project to left- and right-handed fermion chirality as usual. Defining $g_{1,2}^v=  g_R^v \pm g_L^v$, 
we can steer the vectorial and axial couplings via $g_{1,2}^v$.
For the gluon-induced production of the scalar $X$ case in Eq.~\gl{eq:subprocs}, we compute the interaction vertices from
\beeq
{\cal{L}}^{ggX}= -{1\over 4} \left( g_1^s \, G^{\mu\nu}  G_{\mu\nu}\,  X +  g_2^s \, G^{\mu\nu}  \widetilde G_{\mu\nu} \, X \right) \,,
\eeeq
\end{subequations}
Ref.~\cite{Alwall:2007st}, where $\widetilde G_{\mu\nu}$ is the Hodge dual of the non-Abelian $SU(3)$ field strength $G^{\mu\nu}$. For the production of
the tensor particle $X$ from gluons, we again assume the vertex function quoted in Eq.~\gl{eq:tensdec}. This choice corresponds to 
gravitonlike coupling, which, when taken to be universal, is already heavily constrained by Tevatron data 
(see e.g. \cite{Abazov:2010xh} for recent D$\slashed{\rm O}$ searches). 
The $X=2^+$, however, still represents
a valid candidate for our spin and ${\cal{CP}}$ analysis as a state analogous to the composite 
$\Delta^0$ baryon.

\begin{figure*}[t!]
\begin{center}
\includegraphics[width=0.48\textwidth]{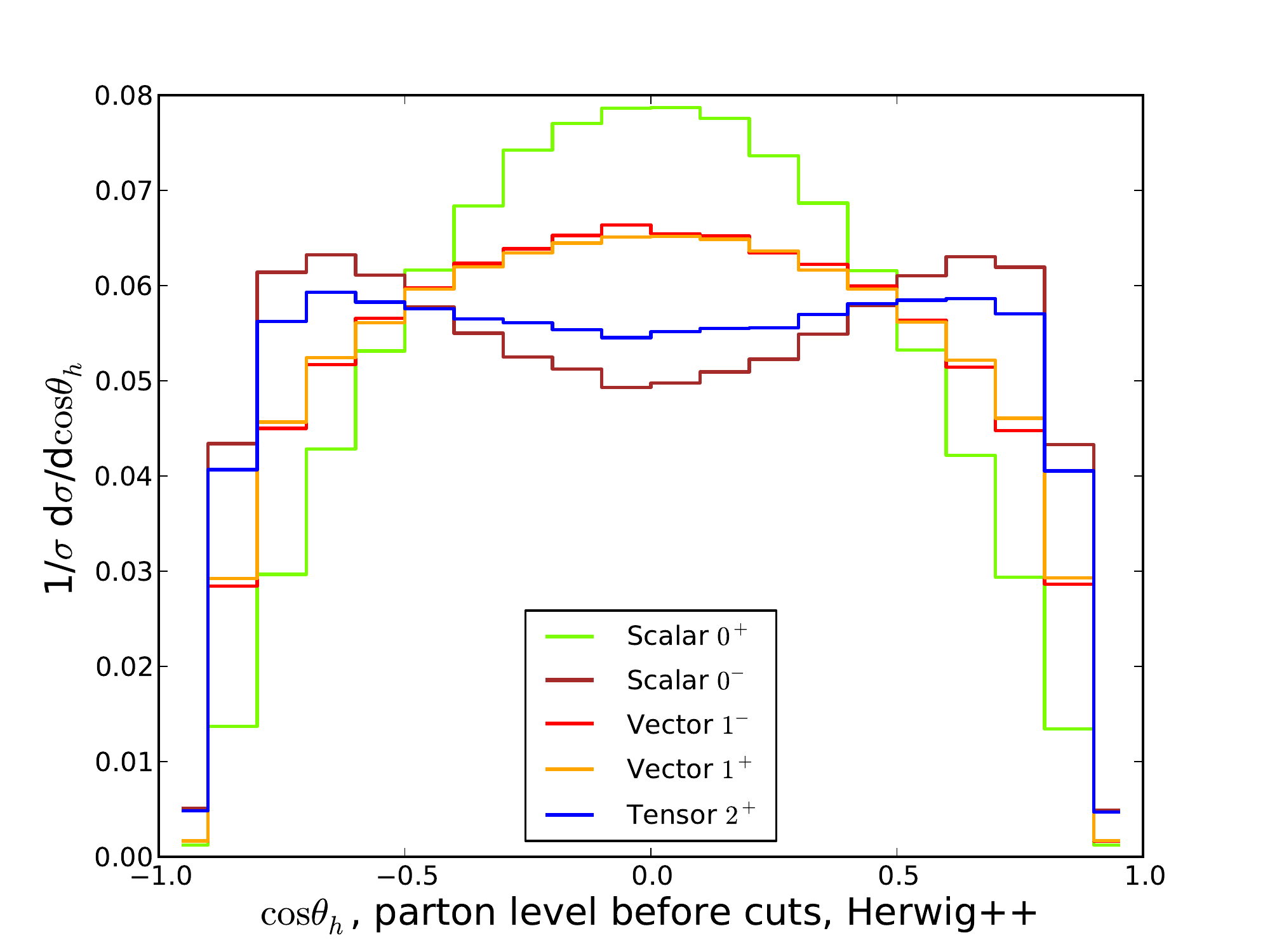}
\hfill
\includegraphics[width=0.48\textwidth]{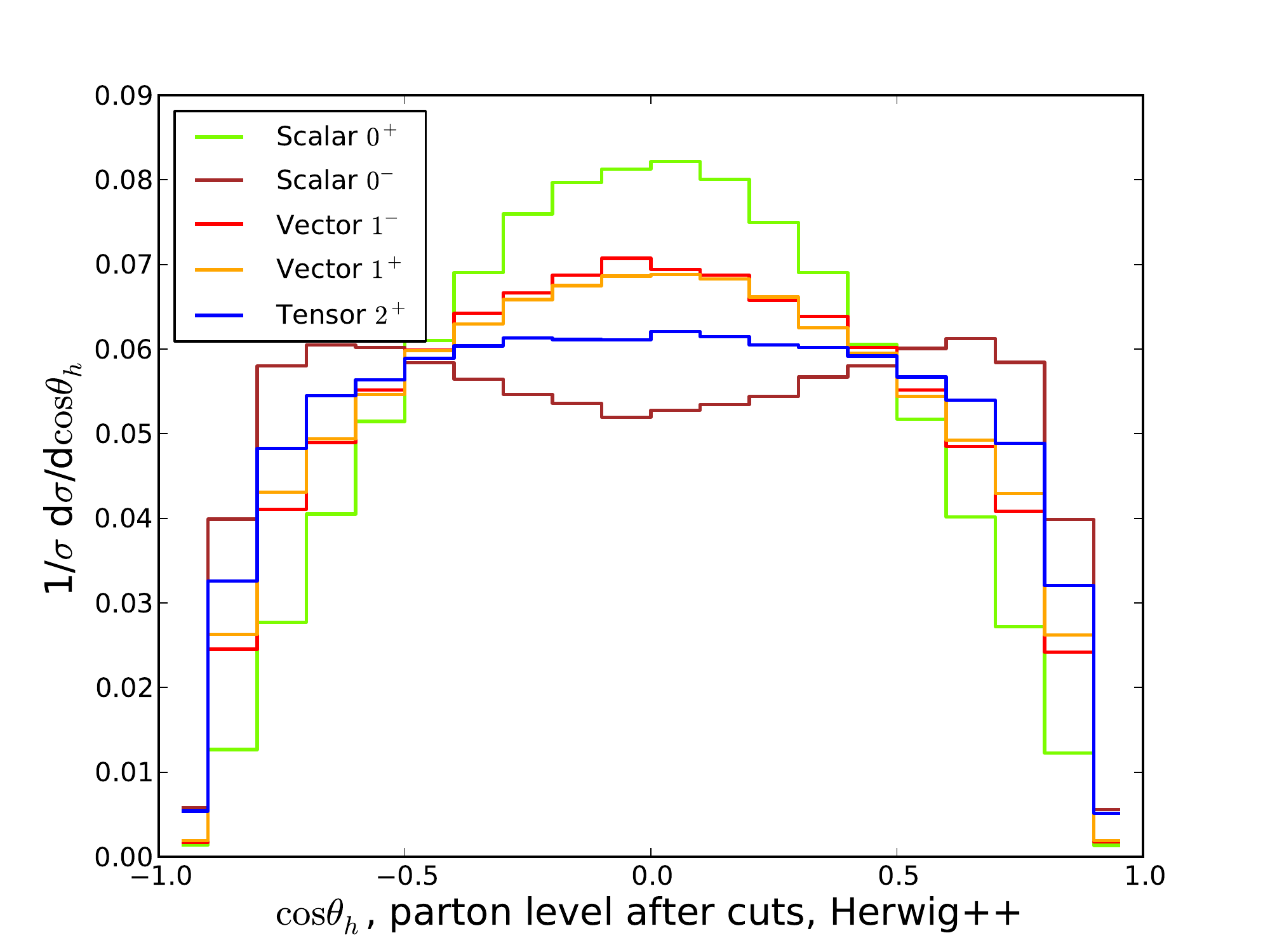}
\includegraphics[width=0.48\textwidth]{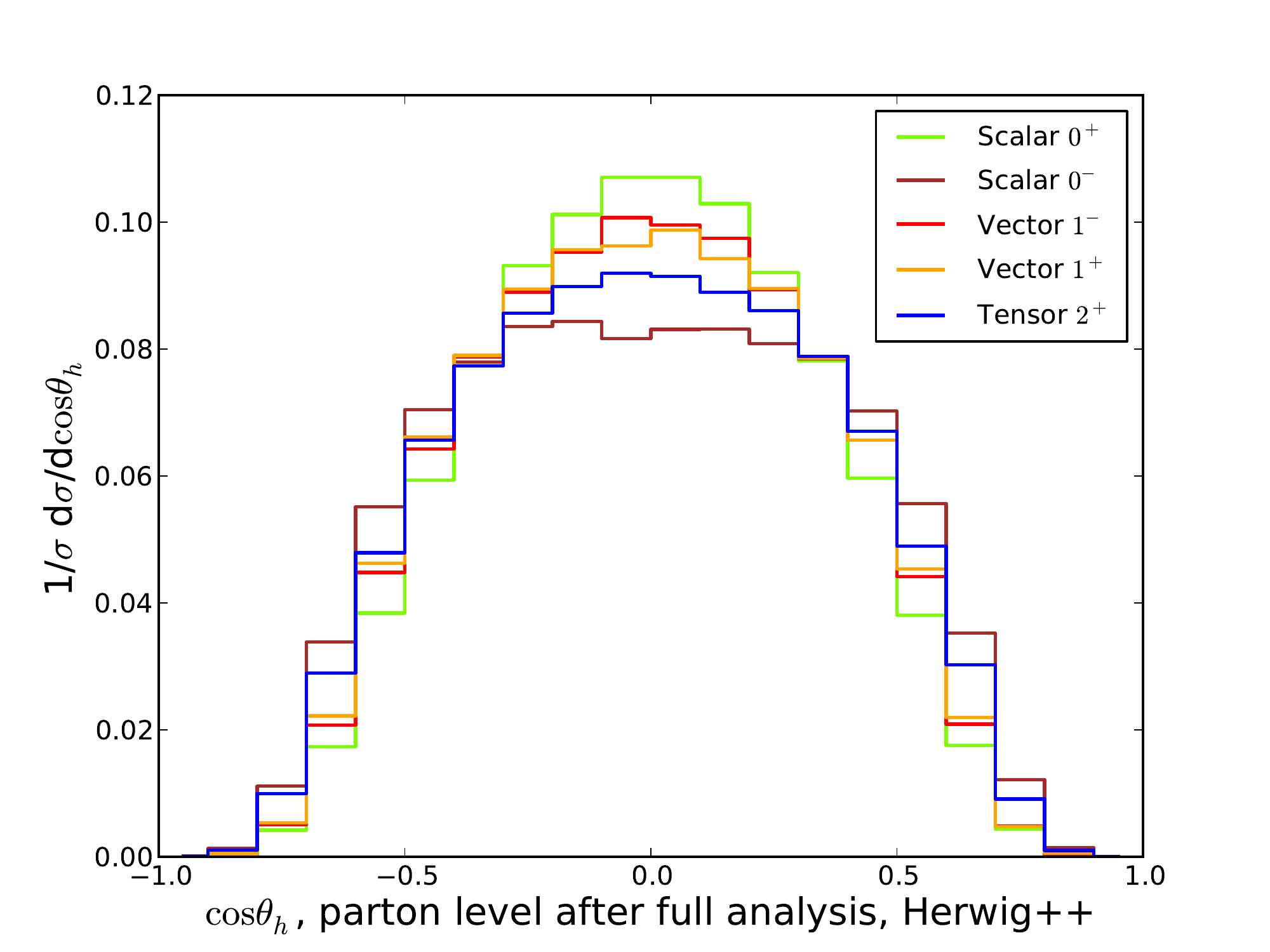}
\hfill
\includegraphics[width=0.48\textwidth]{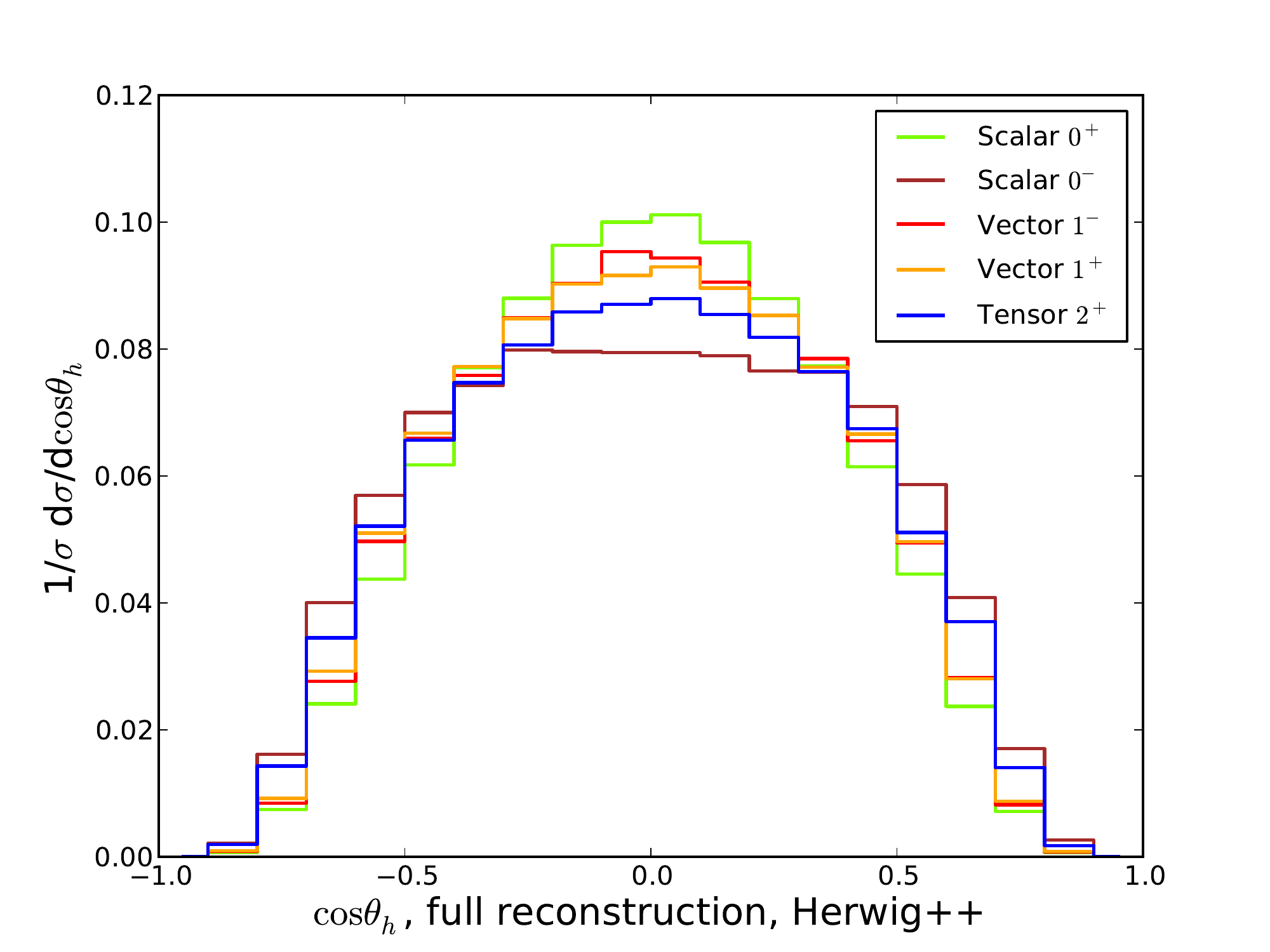}
\end{center}
\caption{\label{fig:helhad} Cosine of the helicity angle $\theta_h$, Eq.~\gl{eq:costhetahel}, calculated from
the hadronically decaying $Z$ at different steps of the analysis:  inclusive Monte Carlo generation level (top, left),
Monte Carlo generation level including selection cuts Eqs.~\gl{eq:mupt}-\gl{eq:fatjetcrit} (top, right), after the full subjet  
analysis including Monte Carlo-truth information (bottom, left), and after the full analysis (bottom, right).}
\vspace{-0.2cm}
\end{figure*}

\begin{figure*}[t!]
\begin{center}
\includegraphics[width=0.48\textwidth]{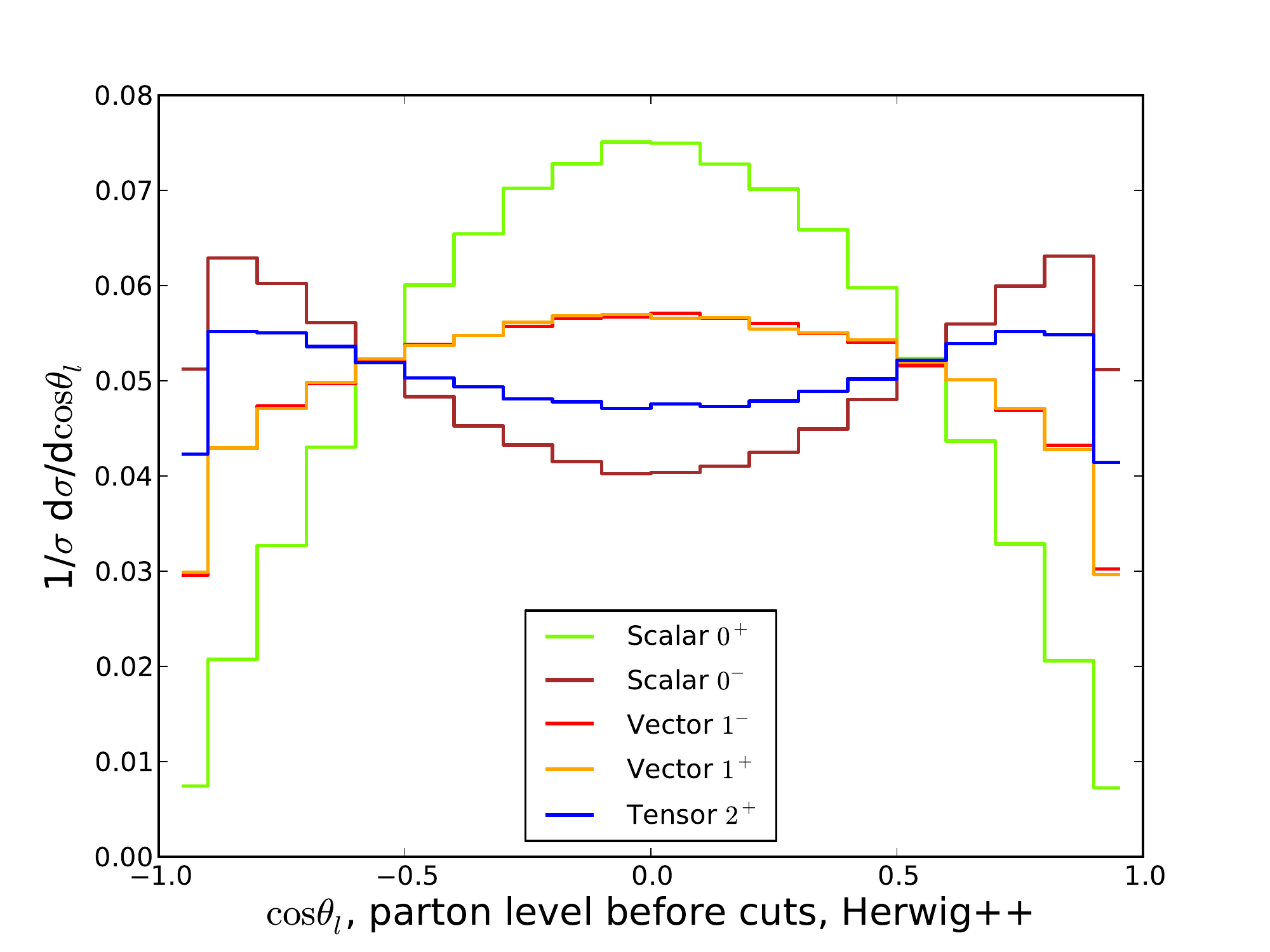}
\hfill
\includegraphics[width=0.48\textwidth]{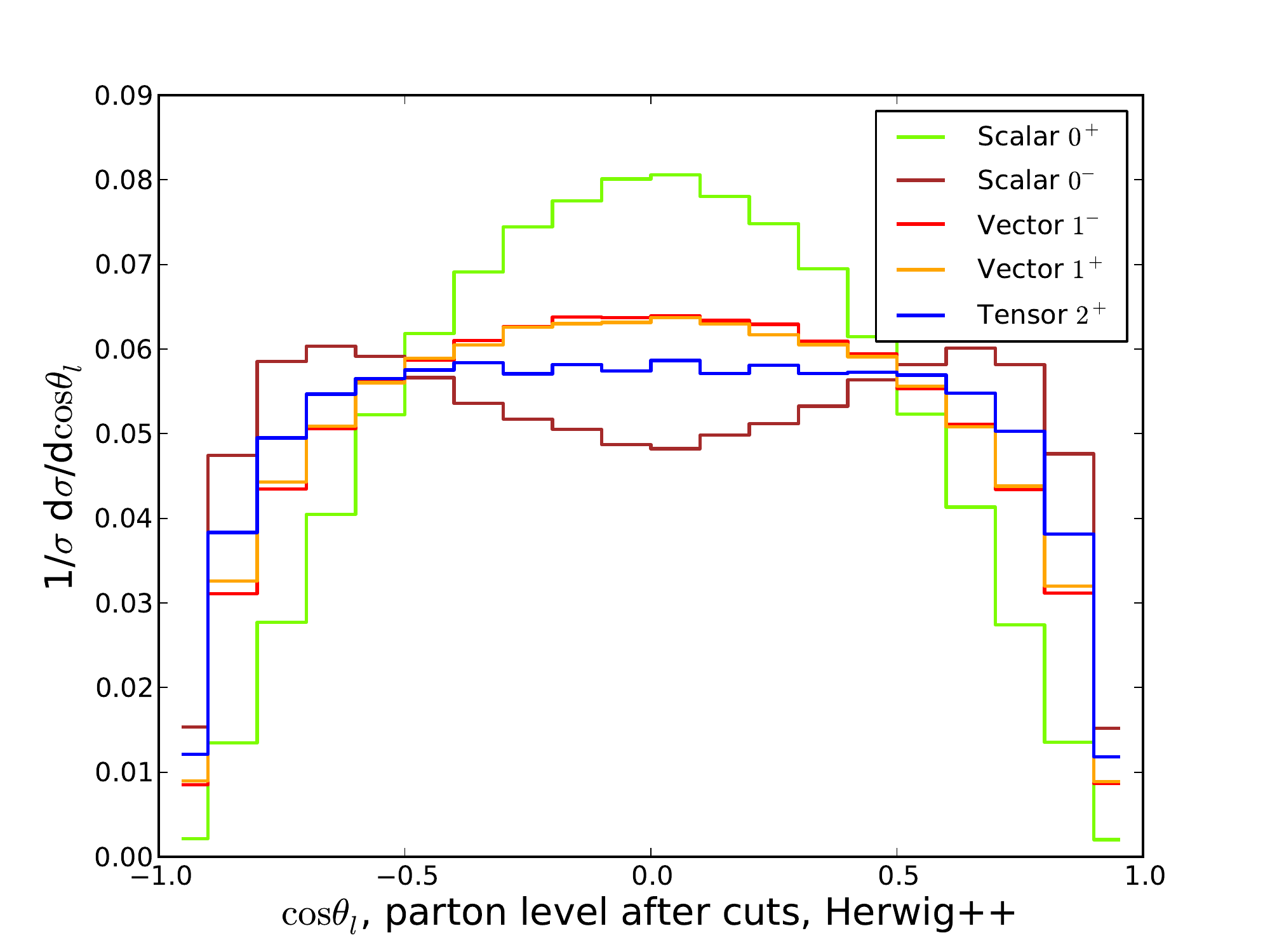}
\includegraphics[width=0.48\textwidth]{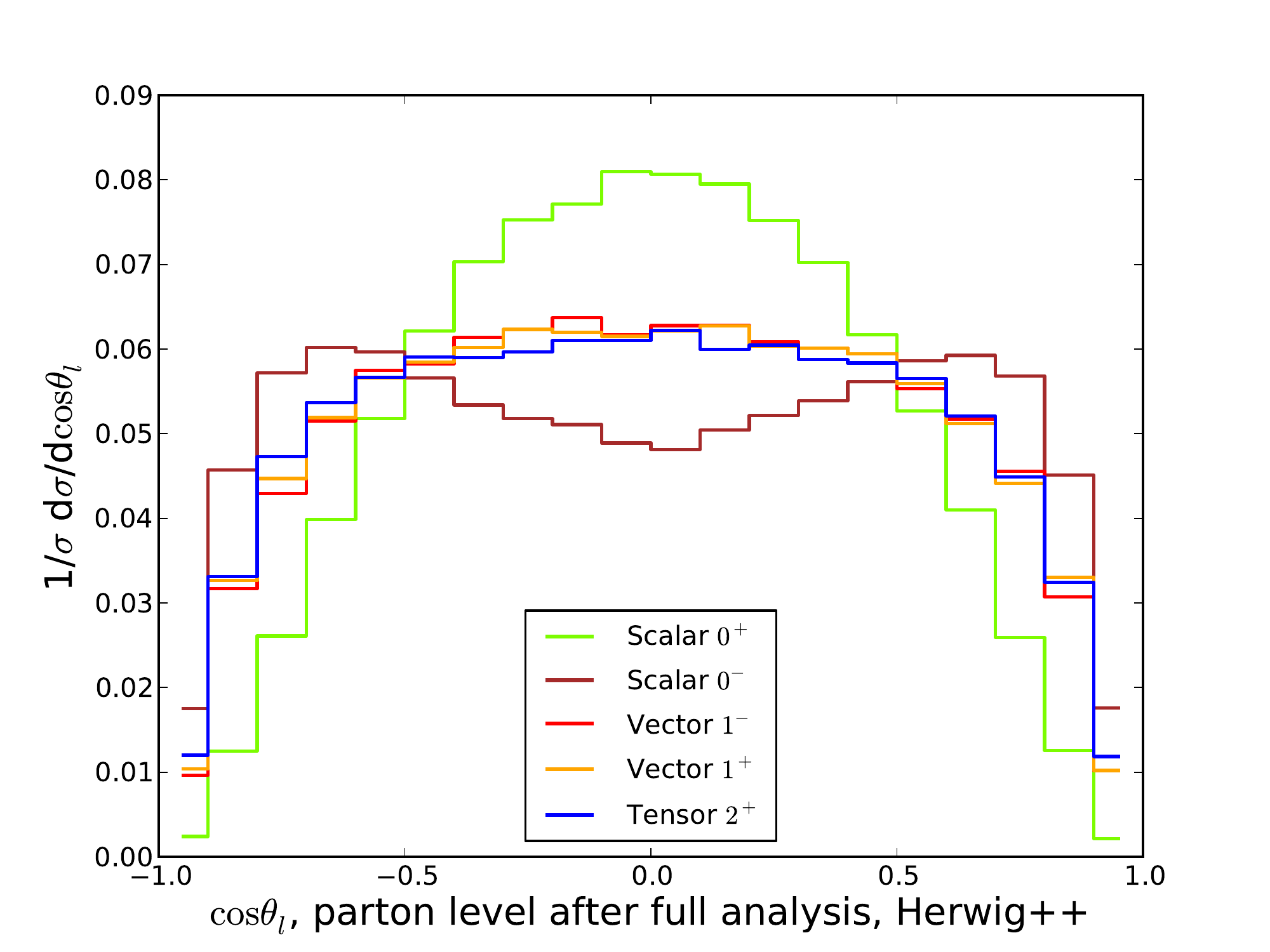}
\hfill
\includegraphics[width=0.48\textwidth]{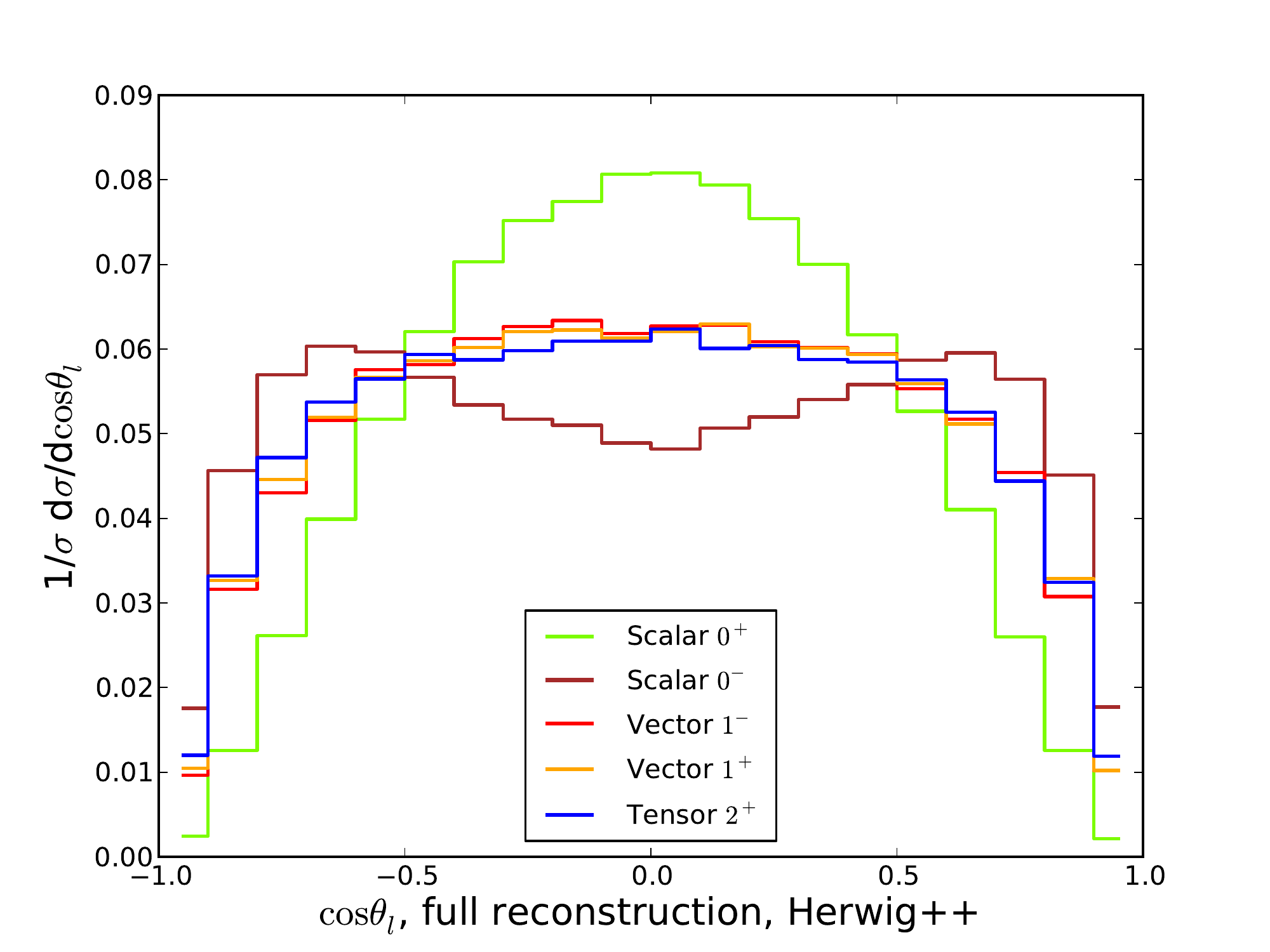}
\end{center}
\caption{\label{fig:hellep} Cosine of the helicity angle $\theta_\ell$, Eq.~\gl{eq:costhetahel}, calculated from
the hadronically decaying $Z$ at different steps of the analysis:  inclusive Monte Carlo generation level (top, left),
Monte Carlo generation level including selection cuts Eqs.~\gl{eq:mupt}-\gl{eq:fatjetcrit} (top, right), after the full subjet  
analysis including Monte Carlo-truth information (bottom, left), and after the full analysis (bottom, right).}
\vspace{-0.2cm}
\end{figure*}

\begin{figure*}[t!]
\begin{center}
\includegraphics[width=0.48\textwidth]{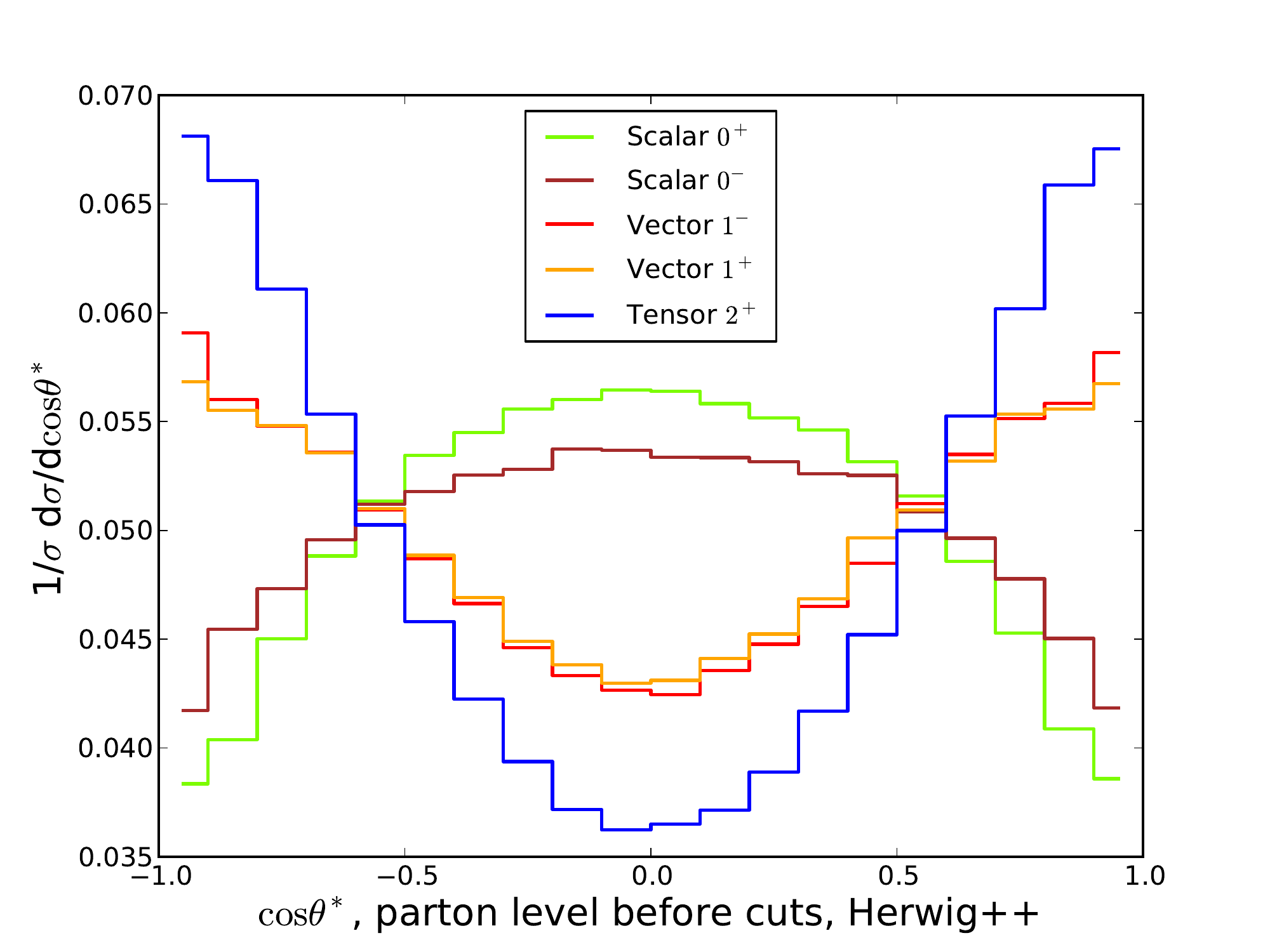}
\hfill
\includegraphics[width=0.48\textwidth]{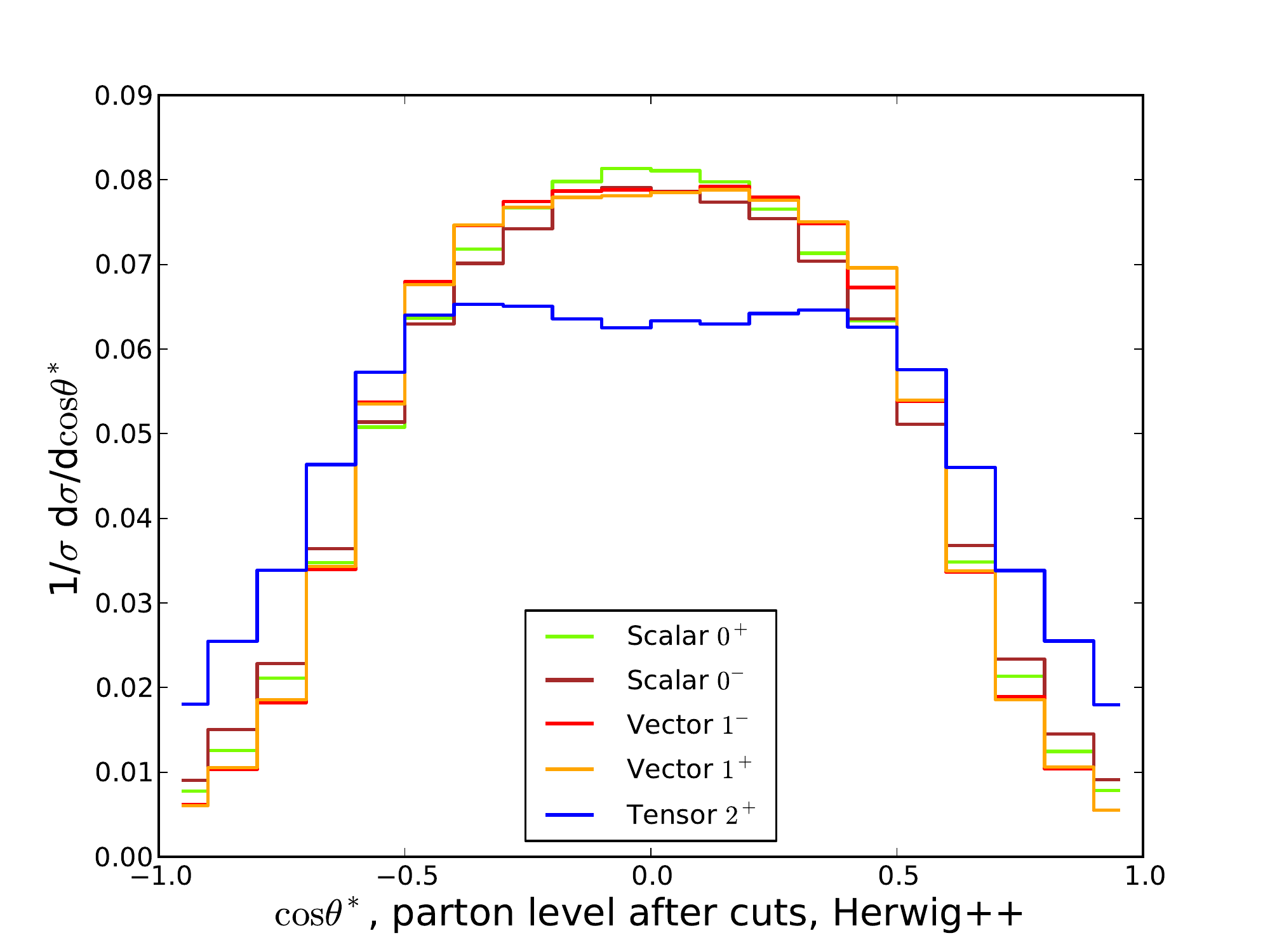}
\includegraphics[width=0.48\textwidth]{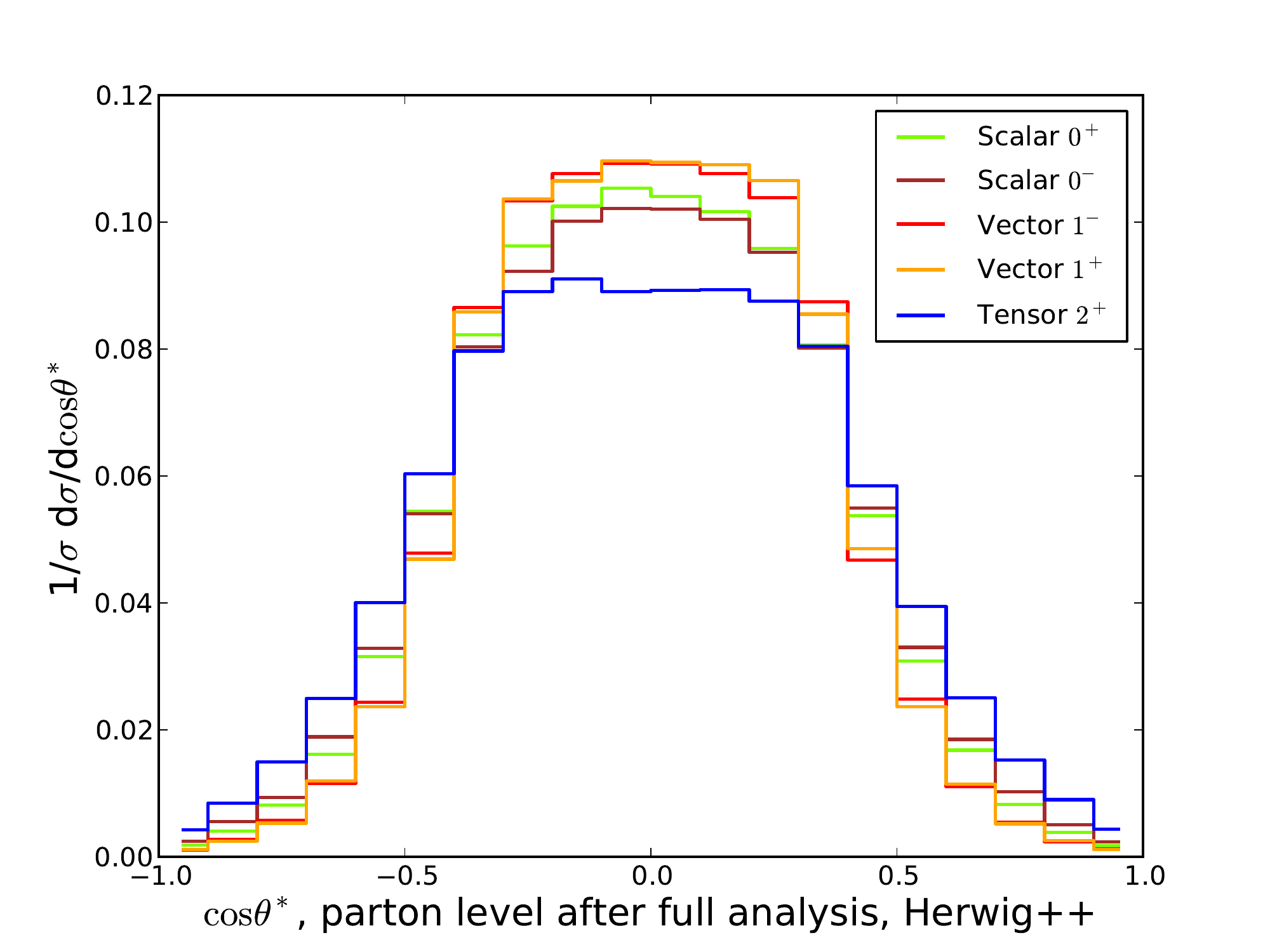}
\hfill
\includegraphics[width=0.48\textwidth]{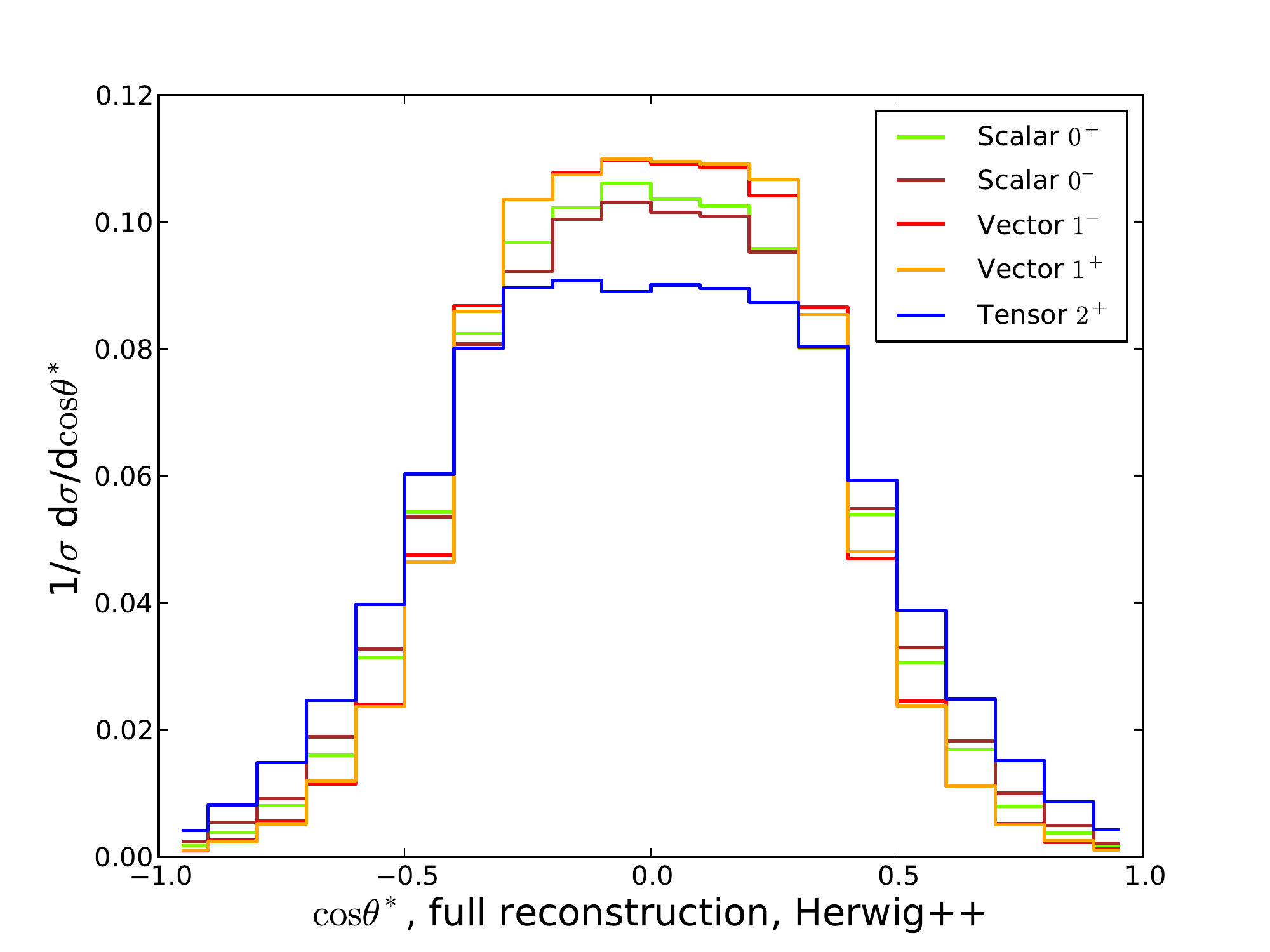}
\end{center}
\caption{\label{fig:anglestar} Cosine of the angle $\theta^\star$, Eq.~\gl{eq:costhetastar}, calculated from
the hadronically decaying $Z$ at different steps of the analysis:  inclusive Monte Carlo generation level (top, left),
Monte Carlo generation level including selection cuts Eqs.~\gl{eq:mupt}-\gl{eq:fatjetcrit} (top, right), after the full subjet  
analysis including Monte Carlo-truth information (bottom, left), and after the full analysis (bottom, right).}
\vspace{-0.2cm}
\end{figure*}

\begin{figure*}[t!]
\begin{center}
\includegraphics[width=0.48\textwidth]{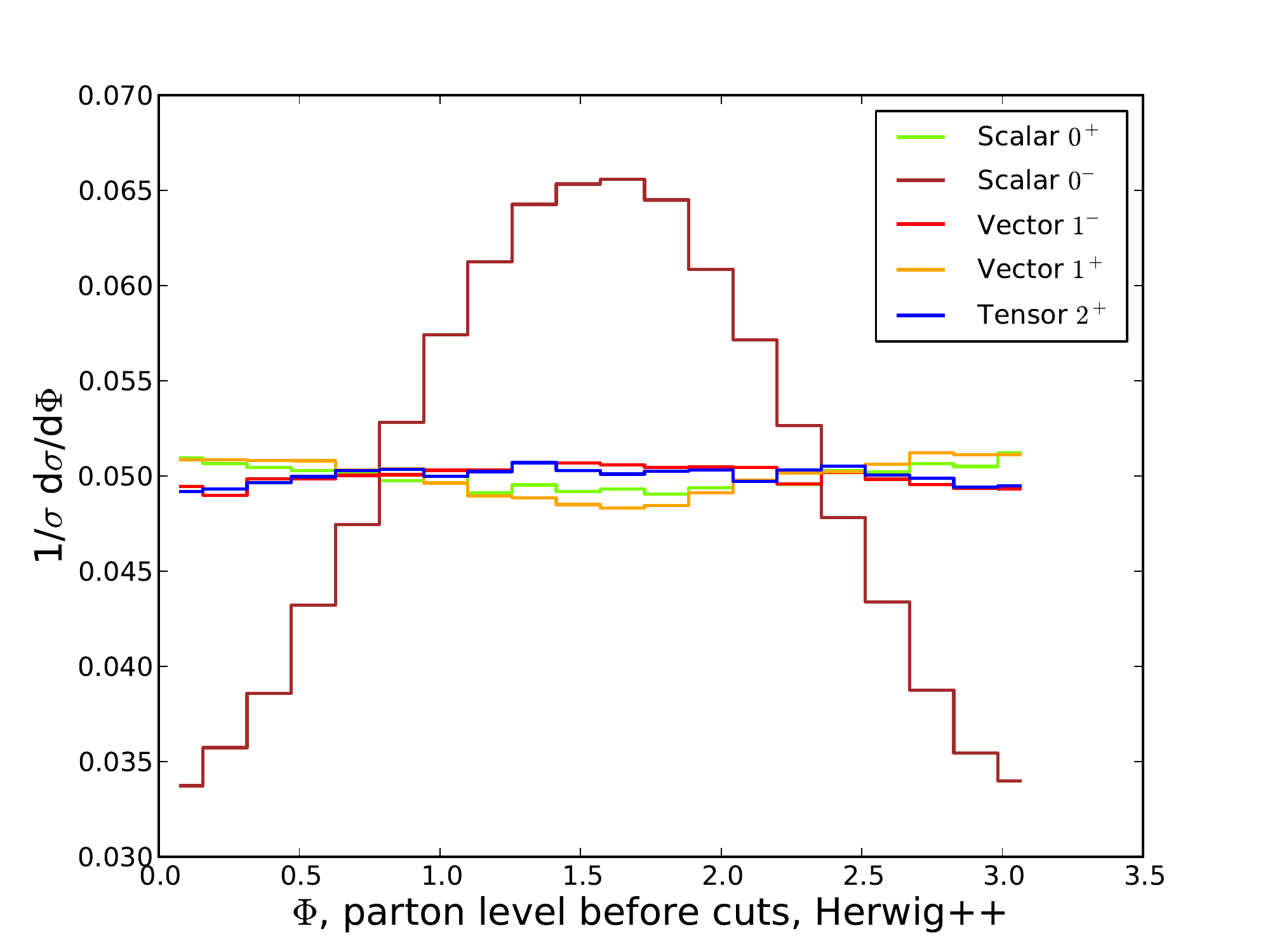}
\hfill
\includegraphics[width=0.48\textwidth]{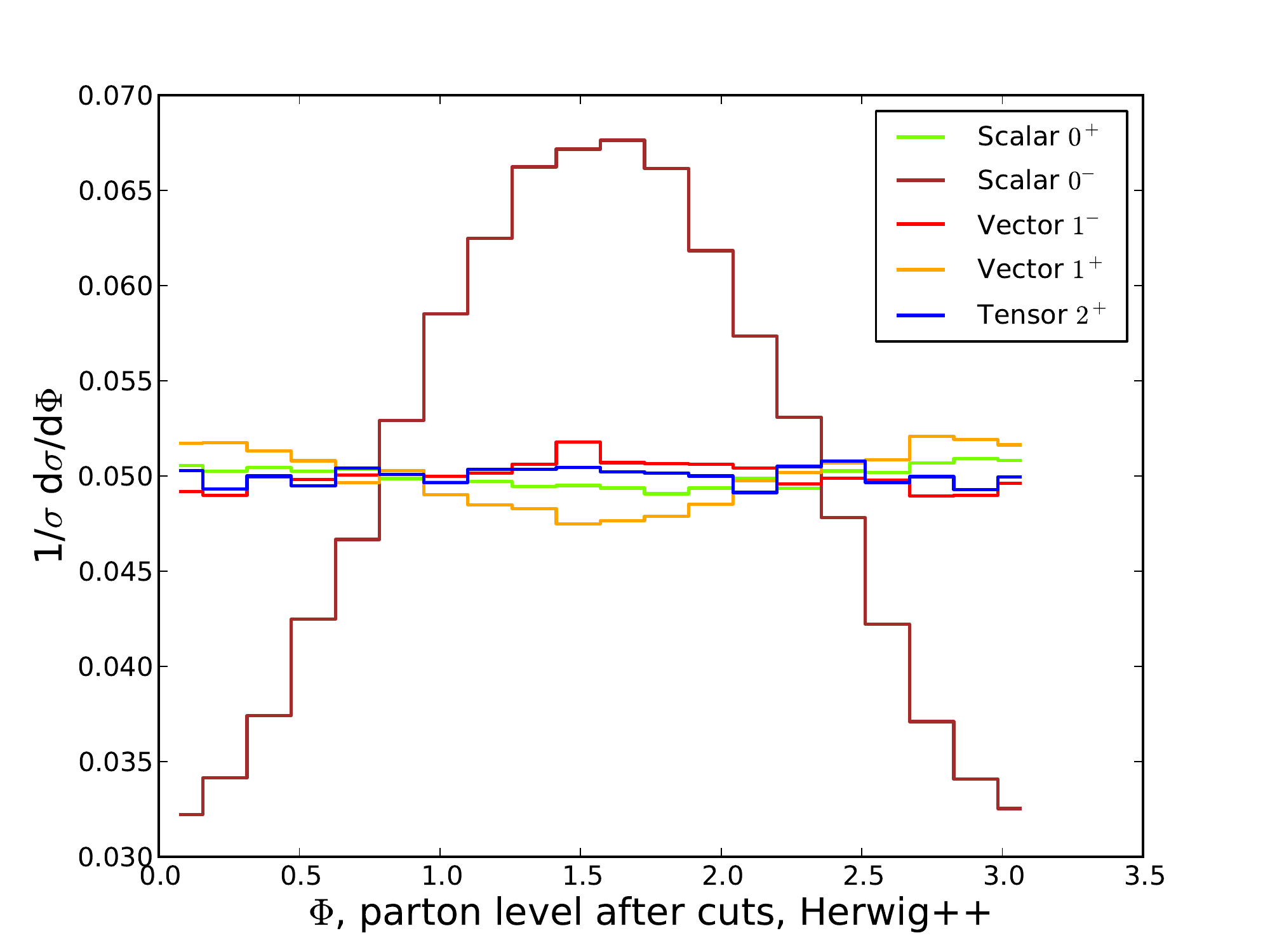}
\includegraphics[width=0.48\textwidth]{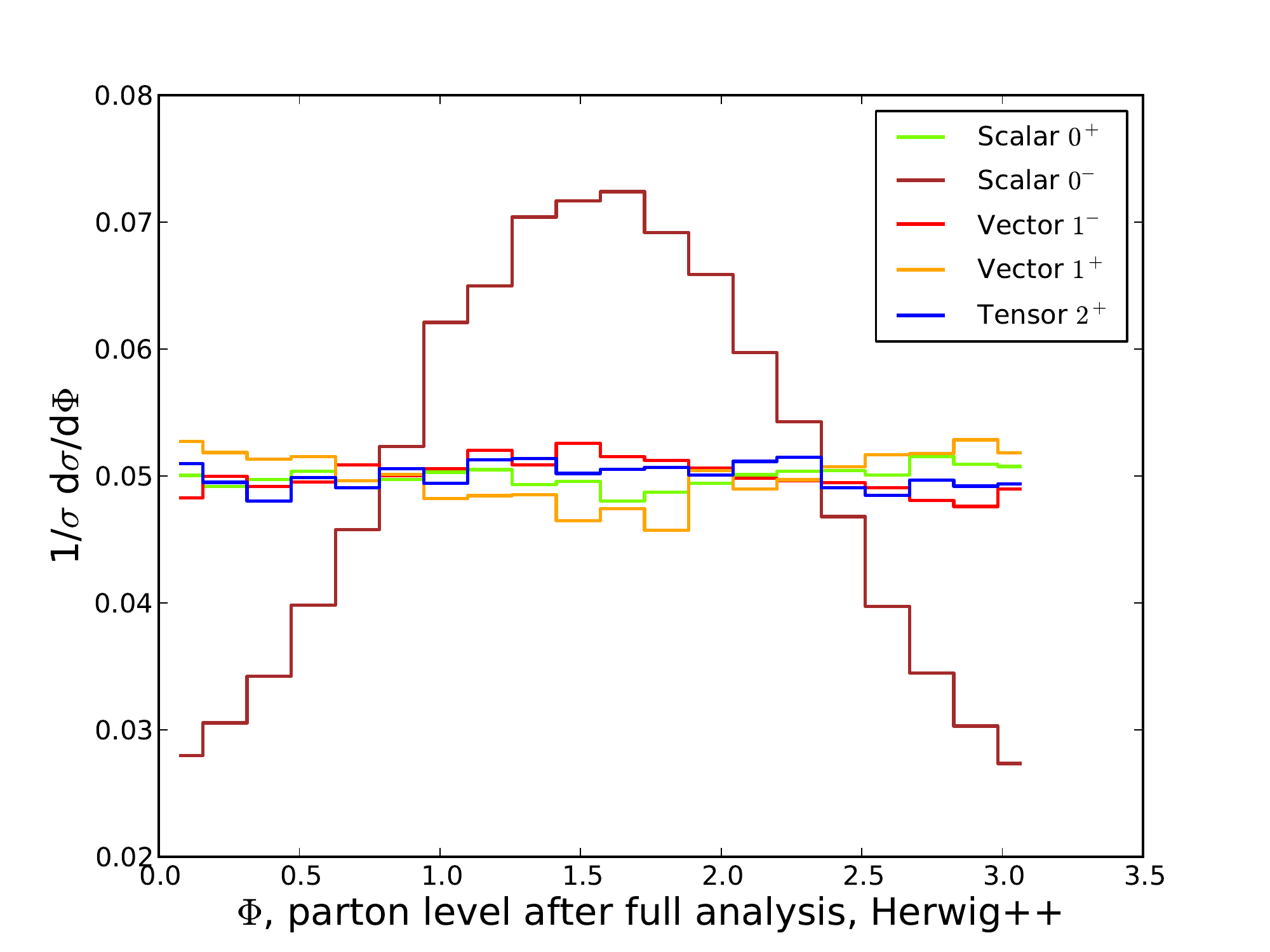}
\hfill
\includegraphics[width=0.48\textwidth]{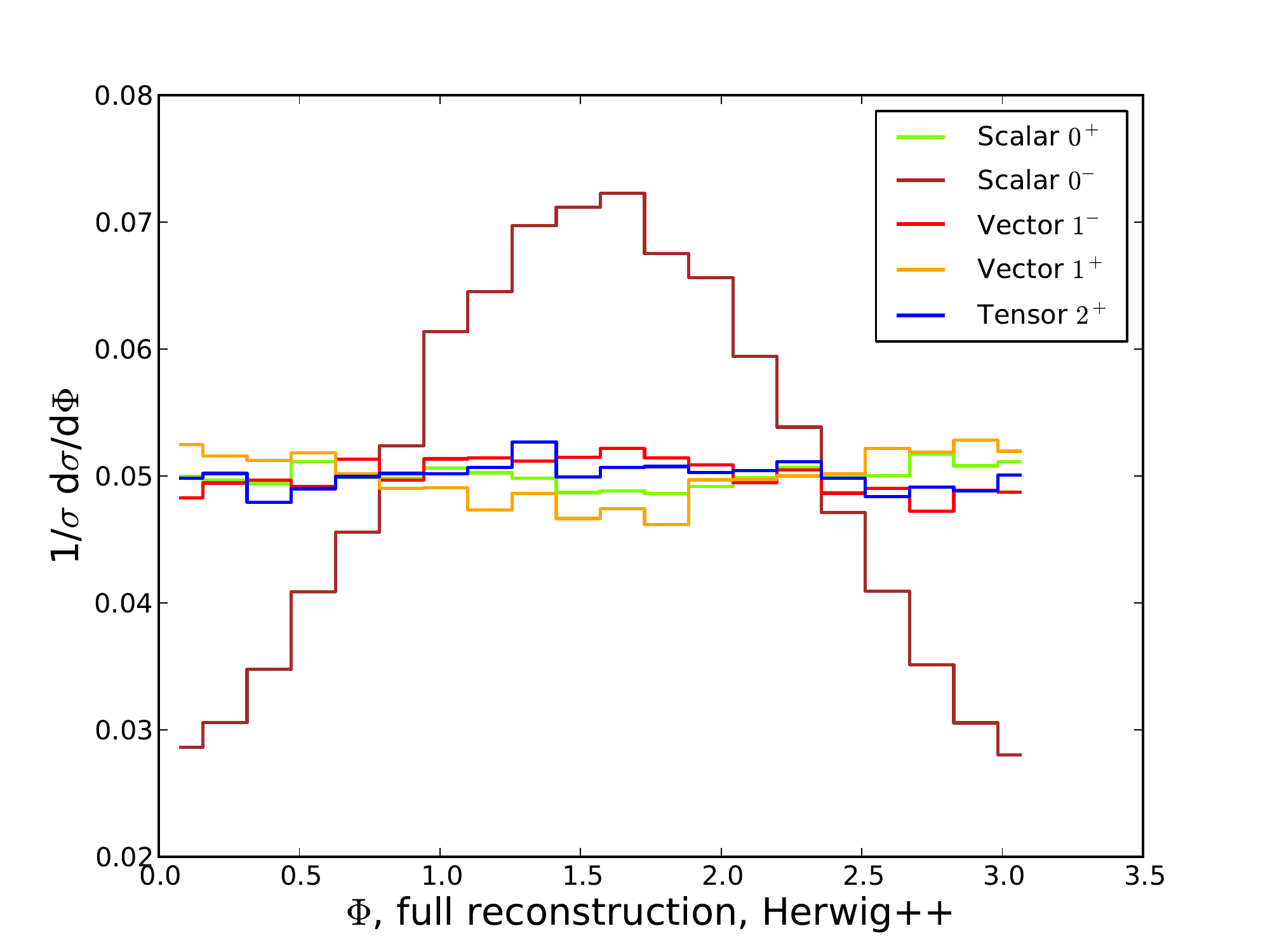}
\end{center}
\caption{\label{fig:phi} Angle $\Phi$, Eq.~\gl{eq:costhetastar}, calculated from
the hadronically decaying $Z$ at different steps of the analysis:  inclusive Monte Carlo generation level (top, left),
Monte Carlo generation level including selection cuts Eqs.~\gl{eq:mupt}-\gl{eq:fatjetcrit} (top, right), after the full subjet  
analysis including Monte Carlo-truth information (bottom, left), and after the full analysis (bottom, right).}
\vspace{-0.2cm}
\end{figure*}

\begin{figure*}[t!]
\begin{center}
\includegraphics[width=0.48\textwidth]{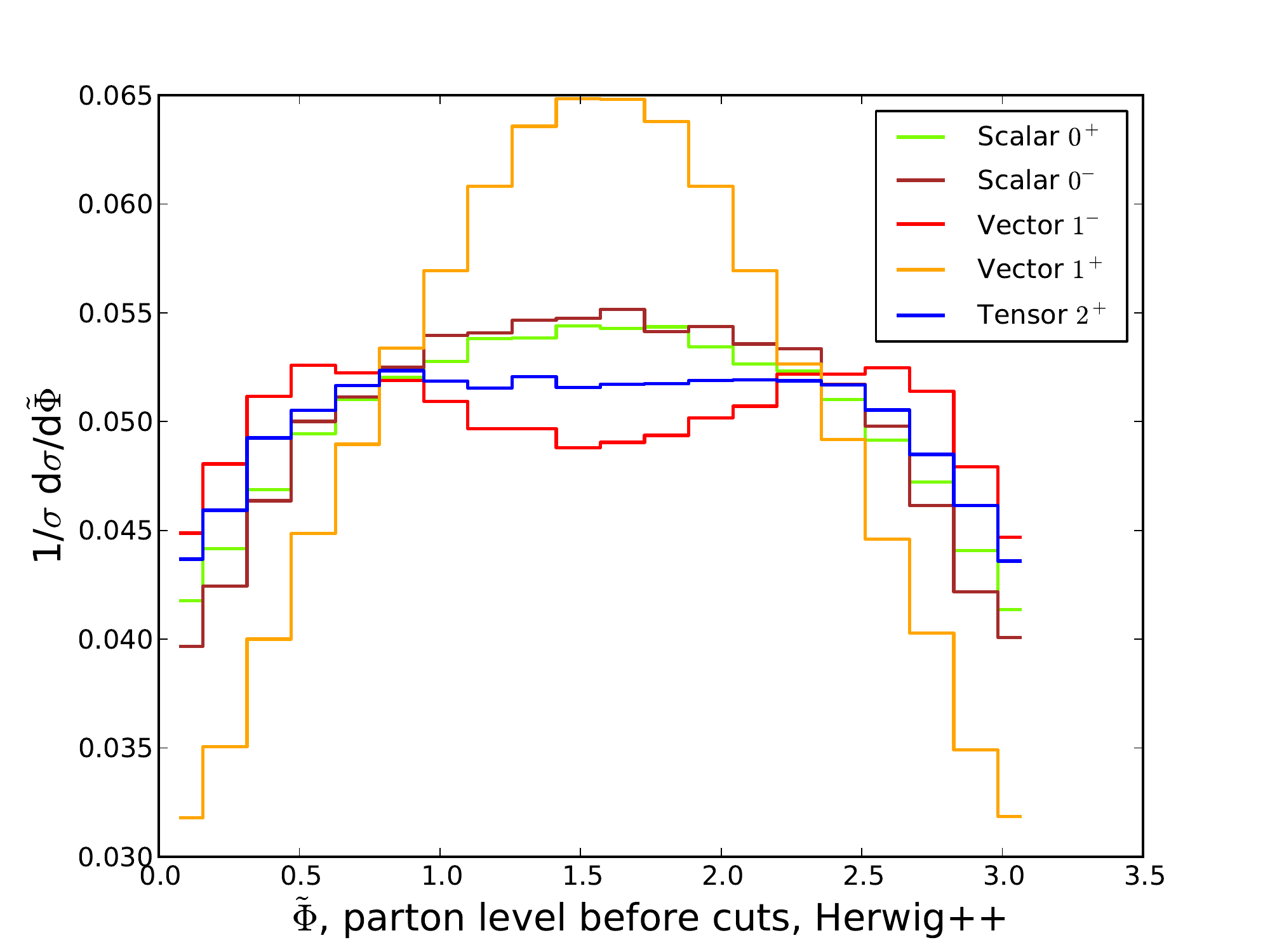}
\hfill
\includegraphics[width=0.48\textwidth]{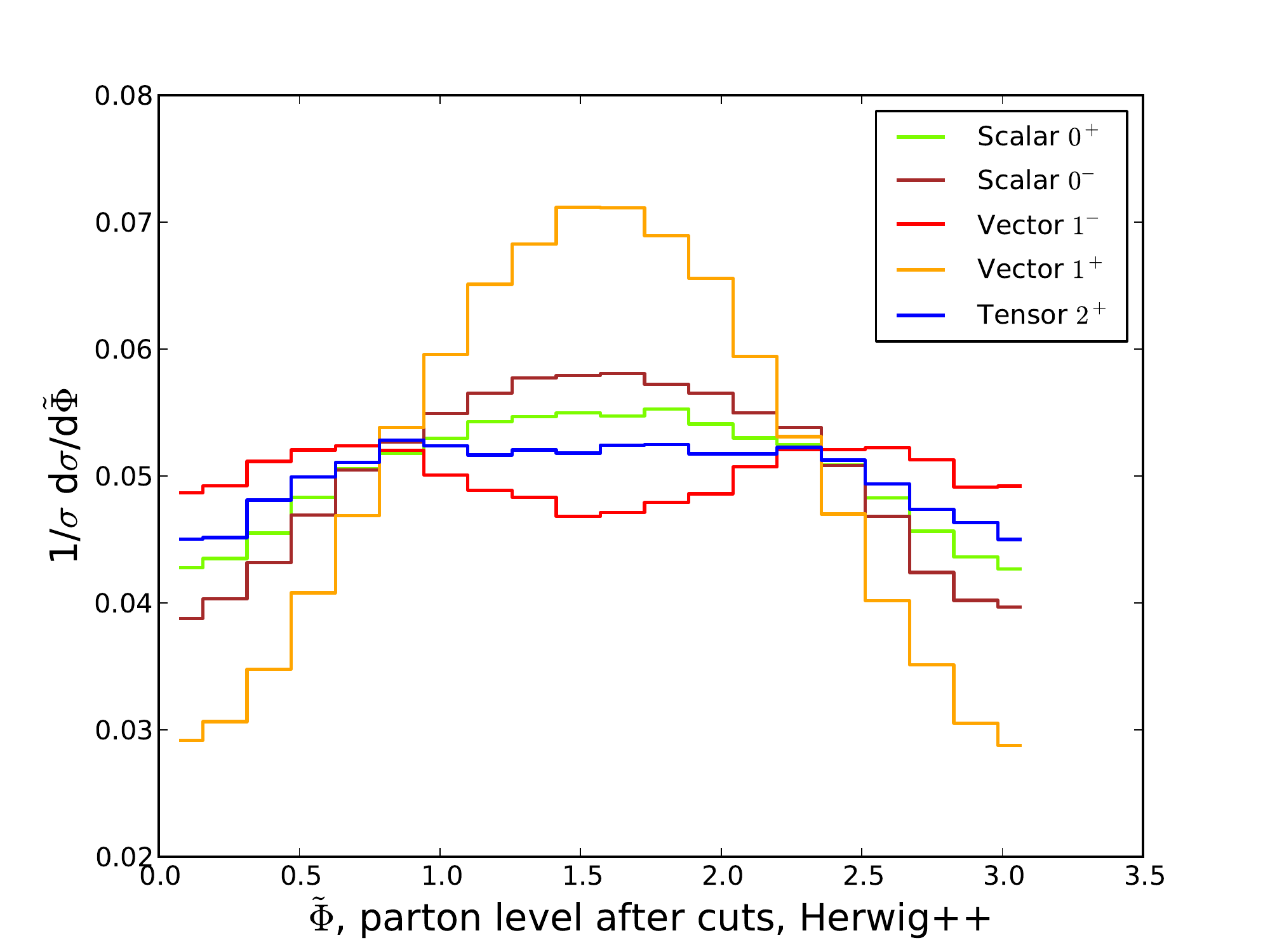}
\includegraphics[width=0.48\textwidth]{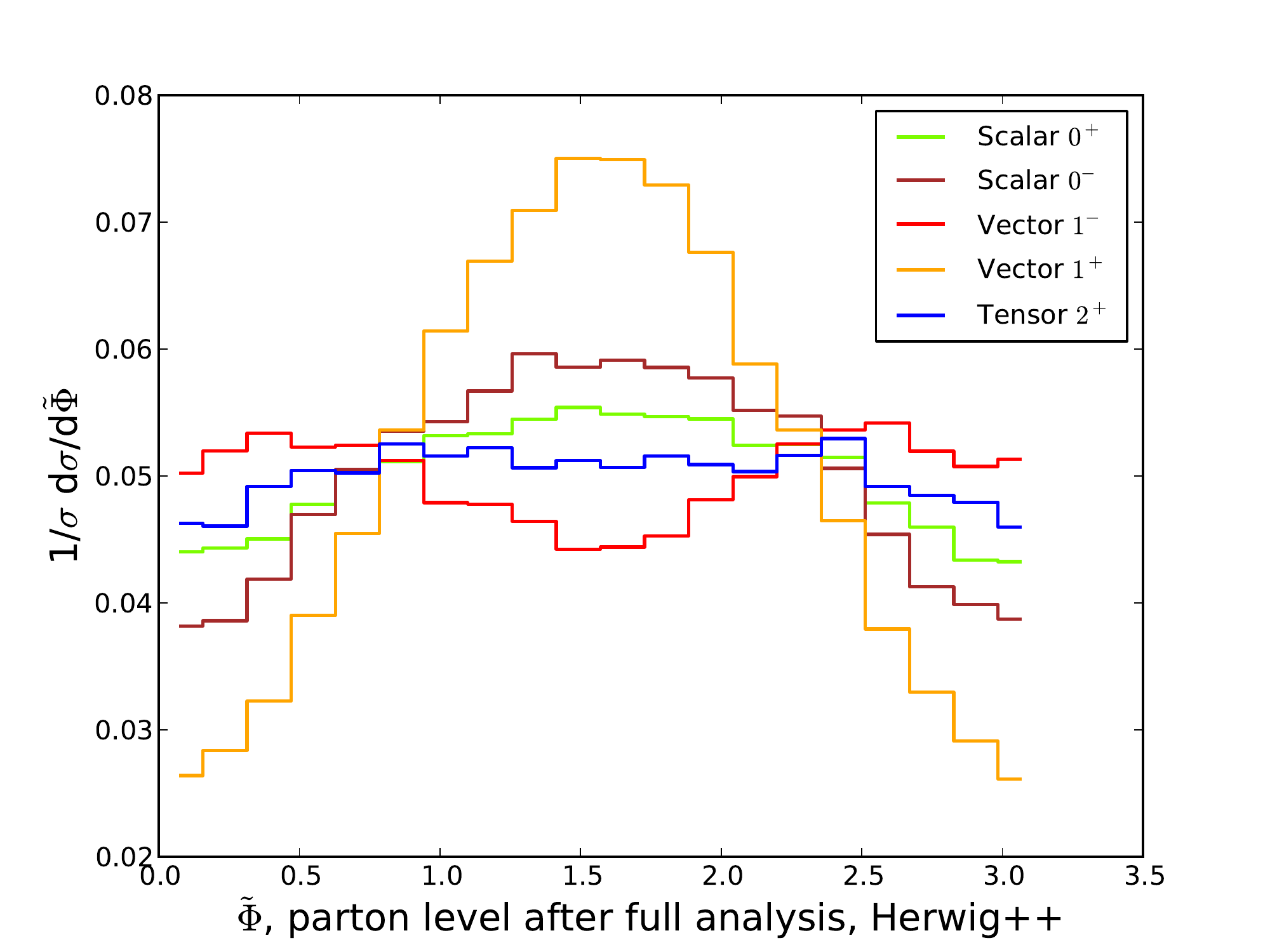}
\hfill
\includegraphics[width=0.48\textwidth]{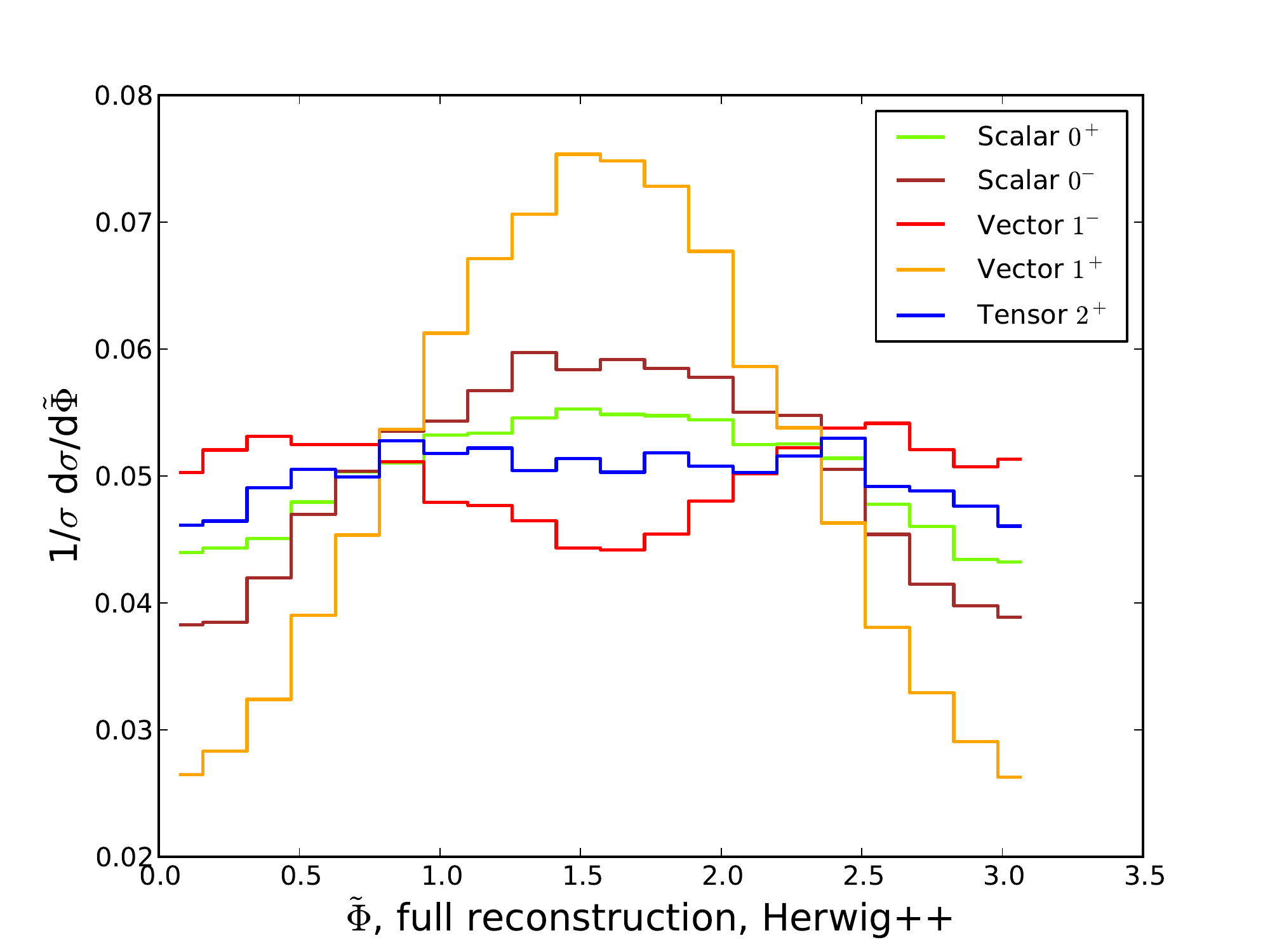}
\end{center}
\caption{\label{fig:phitilde} Angle $\tilde\Phi$, Eq.~\gl{eq:decayplane}, calculated from
the hadronically decaying $Z$ at different steps of the analysis:  inclusive Monte Carlo generation level (top, left),
Monte Carlo generation level including selection cuts Eqs.~\gl{eq:mupt}-\gl{eq:fatjetcrit} (top, right), after the full subjet  
analysis including Monte Carlo-truth information (bottom, left), and after the full analysis (bottom, right).}
\end{figure*}

In the following we consider the five scenarios of Table~\ref{tab:scenarios} for our 
comparison in Sec.~\ref{sec:results} 
for $X$ mass and width choices $m_X=400\gev$ and $\Gamma_X=27\gev$.
The parton level Monte Carlo results for
observables of Eq.~\gl{eq:angledefinition} are plotted in Figs.~\ref{fig:helhad}-\ref{fig:phitilde} of
Sec.~\ref{sec:results}. 
From a purely phenomenological point of view,
our strategy to normalize the parton level cross sections to the SM Higgs
production at next-to-leading order (NLO, $50.84~{\rm{fb}}$ after selection cuts)\footnote{For the NLO Higgs production normalizations we use the 
codes of Refs.~\cite{higlu} and \cite{vbfnlo} for the gluon-fusion and weak-boson-fusion contributions, respectively.}, effectively removes the dependence on
the process-specific combinations of the parameters $c_{1,2}^{s,v},c_{1}^t$ and 
$g_{1,2}^{s,v},g_{1}^t$, as well as the dependence on the initial state 
parton distribution functions on the considered spin- and ${\cal{CP}}$-sensitive
angles. At the same time, the distinct angular correlations will induce different signal efficiencies for
the different particles $X=J^{{\cal{CP}}}$, when the signal sample is 
confronted with the subjet analysis' selection cuts. In our approach, these naturally communicate to the final state after 
showering \mbox{and~hadronization.}

\begin{table}[b!]
\setcounter{table}{0}
\begin{tabular}[b]{c c c}
\hline
$J^{\cal{CP}}(X)$ & Production Eq.~\gl{eq:proddens} & Decay Eq.~\gl{eq:decdens} \\
\hline
$0^+$ & $g_1^s\neq 0,~g_2^s=0 $ &  $c_1^s\neq 0,~c_2^s=0$ \\
$0^-$ & $g_1^s= 0,~g_2^s\neq 0 $ &  $c_1^s=0,~c_2^s\neq 0$\\
$1^+$ & $g_1^v= 0,~g_2^v\neq 0 $ &  $c_1^v=0,~c_2^v\neq 0$\\
$1^-$ & $g_1^v\neq 0,~g_2^v= 0 $ &  $c_1^v\neq 0,~c_2^v= 0$\\
$2^+$& $g_1^t\neq 0$ & $c_1^t\neq 0$  \\
\hline
\end{tabular}
\caption{\label{tab:scenarios} Definition of the scenarios considered for the comparison
in Sec.~\ref{sec:results} for the $X$ mass and width choices $m_X=400\gev$ and $\Gamma_X=27\gev$.}
\end{table}

In principle, the $s$-channel signal adds coherently to the continuum $ZZ$ production
and their subsequent decay. We consider these Feynman graphs as part of the background and
discard the resulting interference terms, which is admissible in the vicinity of the resonance. We have explicitly
checked the effect of the interference on the angular distributions at the parton level for inclusive generator-level cuts
and find excellent agreement for the invariant $X$ mass window around the resonance,
which is later applied as a selection cut in the subjet analysis.

We further process the {\sc{MadEvent}}-generated signal events
with \textsc{Herwig++} \cite{Bahr:2008pv} for parton showering and hadronization. \textsc{Herwig++} 
includes spin correlations to the shower, hence minimizing unphysical contamination
of the backgrounds' angular distribution by simulation-related shortcomings. 
We have also compared the results to {\textsc{Pythia}} 6.4 \cite{Sjostrand:2006za} 
to assess the systematic uncertainties and find reasonable agreement for the net efficiencies 
after all analysis steps have been carried out 
(see Table~\ref{tab:cutscenarios} and the discussion of the next section).

%--------------------------------------------------------------------------------------------------------------------------------------------------------
%--------------------------------------------------------------------------------------------------------------------------------------------------------
% Helicity Angle Had.
%--------------------------------------------------------------------------------------------------------------------------------------------------------
%--------------------------------------------------------------------------------------------------------------------------------------------------------
% Helicity Angle Lep.
%--------------------------------------------------------------------------------------------------------------------------------------------------------
%--------------------------------------------------------------------------------------------------------------------------------------------------------
% Helicity AngleStar
%--------------------------------------------------------------------------------------------------------------------------------------------------------
%--------------------------------------------------------------------------------------------------------------------------------------------------------
% Phi
%--------------------------------------------------------------------------------------------------------------------------------------------------------
%--------------------------------------------------------------------------------------------------------------------------------------------------------
%  TildePhi
%--------------------------------------------------------------------------------------------------------------------------------------------------------
\begin{figure*}[t!]
\begin{center}
\includegraphics[width=0.48\textwidth]{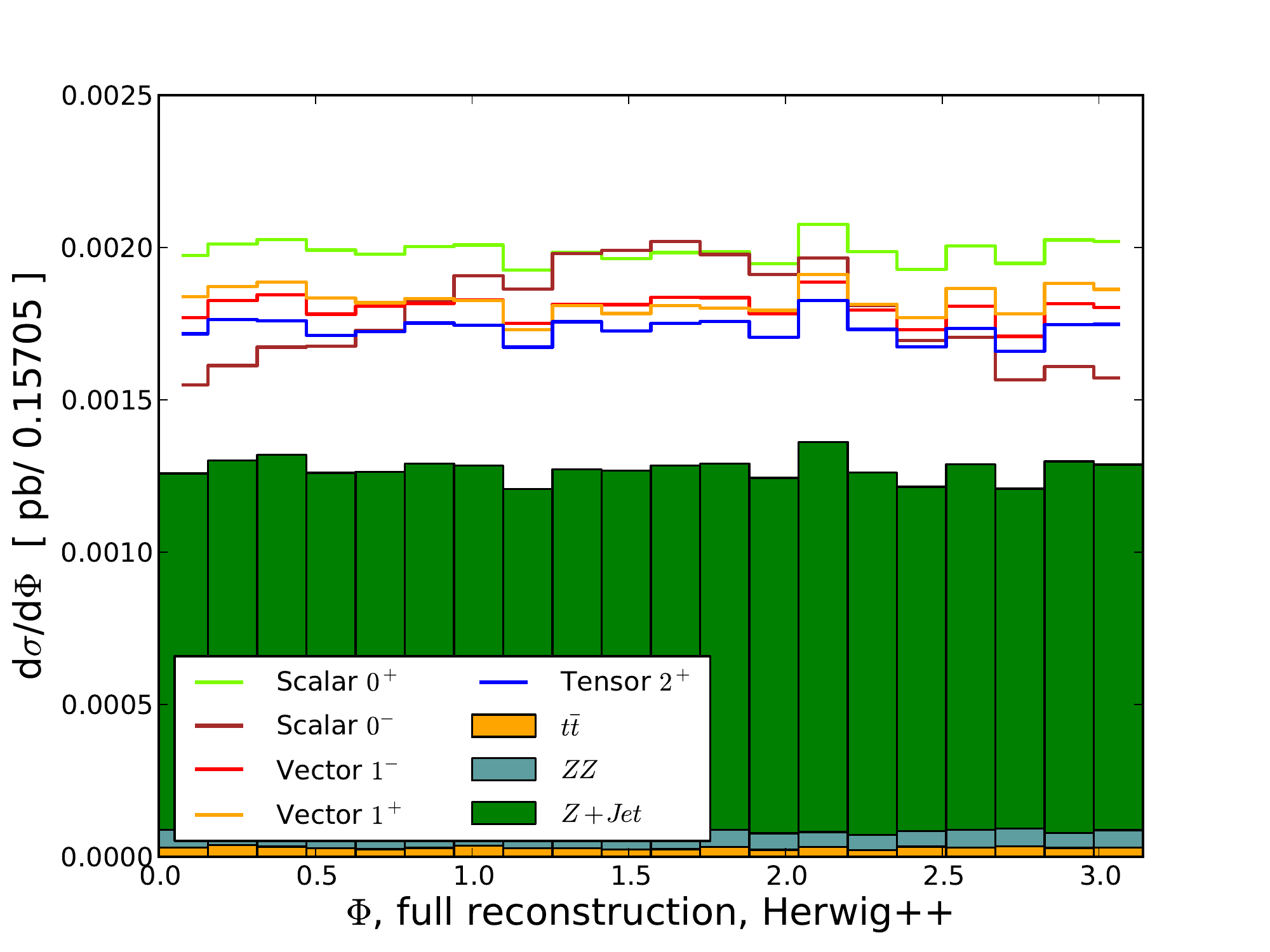}
\\
\includegraphics[width=0.48\textwidth]{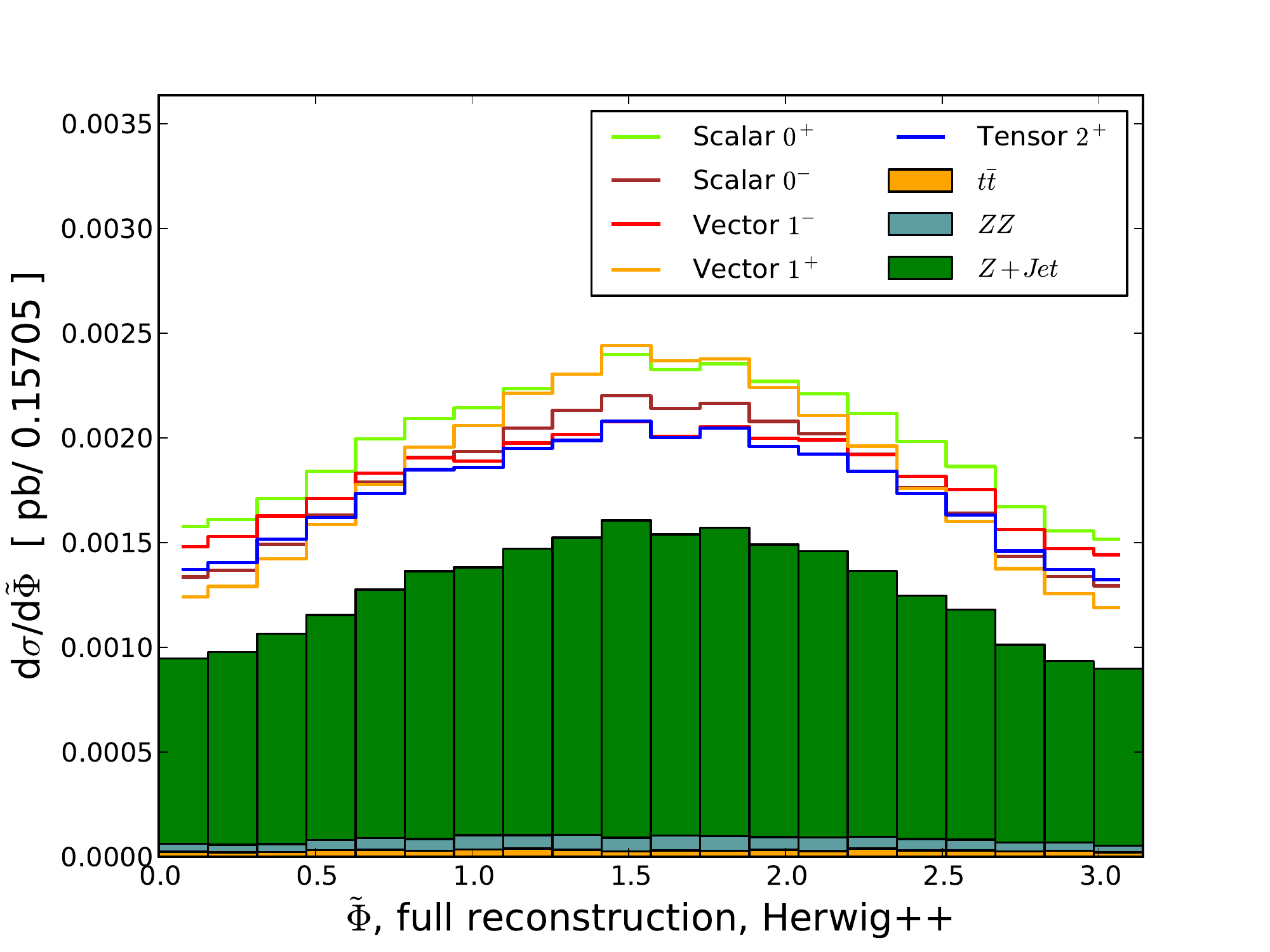}
\hfill
\includegraphics[width=0.48\textwidth]{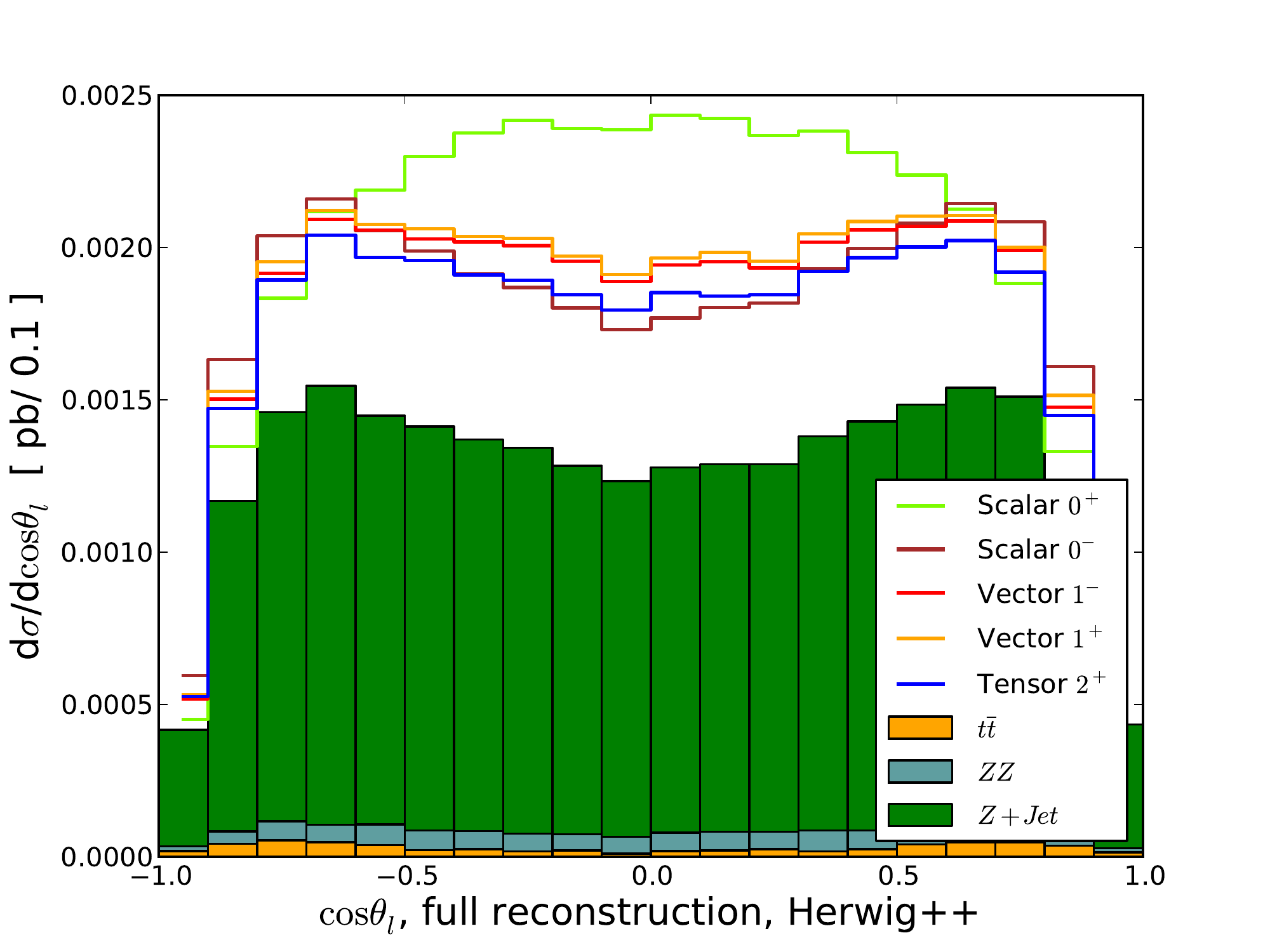}
\end{center}
\caption{\label{fig:anglebkg}The spin- and $\cal{CP}$-sensitive angles $\Phi$ (top), $\tilde\Phi$ (bottom, left) and 
$\cos\theta_\ell$ (bottom, right) including the shape of the backgrounds, simulated with {\sc{Herwig++}}.}
\vspace{-0.2cm}
\end{figure*}

We include the dominating background processes to our analysis. These are
$Z+$jets, $ZZ$, $t \bar{t}$ and $WZ$ production. For the main 
background $Z+\text{jets}$ the NLO QCD cross section, 
requiring $p_T({\rm{jet}}) \geq 100\gev$, is $33.9$ pb~\cite{mcfm}. 
The NLO QCD normalization of the $t\bar{t}$ production cross section is $875$ pb 
\cite{Cacciari:2008zb}, and for $WZ$ production we find  $43.4$ pb \cite{mcfm}. 
The NLO QCD $ZZ$ background is $19.0$ pb. For the simulation of the SM Higgs boson we take the NLO gluon-fusion and weak-boson-fusion production mechanism into account and normalize all signals to the inclusive production cross section. 
We simulate the backgrounds
with \textsc{MadEvent}, again employing \textsc{Pythia} 6.4 and {\sc{Herwig++}} for showering, 
and normalize the distributions to the NLO QCD cross section approximation.
For the weak-boson-fusion channels, the NLO QCD corrections are known
to be at the percent level for a deep inelastic scattering type of factorization scale choices (see
Ref.~\cite{wbfnlo,vbfnlo}). For these processes, the electroweak modifications then become important since they
turn out to be comparable in size \cite{Ciccolini:2007jr}. As we perform simulations
within an effective electroweak theory approach, and for the smallness of the overall corrections,
compared to additional uncertainties presently inherent to showering \cite{Alwall:2007fs},
we set $\sigma^{\rm{NLO}}/\sigma^{\rm{LO}}=1$ for the weak-boson-fusion contributions.
It turns out that the $WZ$ background is negligibly small, and we therefore only focus
on the $Z+{\rm{jets}}$, $ZZ$, and $t\bar t$ backgrounds in the discussion of our results 
in~Sec.~\ref{sec:results}. 

From the fully-simulated Monte Carlo event, we reconstruct
the detector calorimeter entries by grouping all final state particles into
cells of size $\Delta \eta\times \Delta \phi = 0.1 \times 0.1$ in the pseudorapidity--azimuthal angle plane, 
to account for finite resolution of the calorimetry. 
The resulting cells' three-momenta are subsequently rescaled such to yield massless cell entries 
\mbox{$p^\mu p_\mu=0$} (see, e.g., Ref.~\cite{Ball:2007zza}).
For the rest of the analysis we discard cells with energy-entries below a calorimeter
threshold of $0.5~\text{GeV}$.

\subsection{Discriminating signal from background}
\label{sec:discr}
To separate the signal events of Eq.~\gl{eq:subprocs} from the background, we perform a fat jet/subjet
analysis. The technical layout has been thoroughly discussed in the recent literature, 
e.g. in Refs.~\cite{Butterworth:2008iy,Plehn:2009rk,Soper:2010xk,Hackstein:2010wk}.
We focus on the following on muon production.

Prior to the jet analysis, we therefore require two isolated muons with 
\beeq
\label{eq:mupt}
p_T (\mu^\pm)>15 \gev \quad \hbox{and} \quad |\eta (\mu^\pm)| < 2.5
\eeeq
in the final state, which reconstruct the invariant $Z$ mass up to $10\gev$,
\beeq
\label{eq:muptz}
\left[p(\mu^+)+p(\mu^-)\right]^2 = (m_Z\pm 10\gev )^2\,.
\eeeq
To call a muon isolated we require that $E_{T_{\rm{had}}} < 0.1\, E_{T_{\mu^{\pm}}}$ within a cone of $R=0.3$ around the muon.
We 
ask for a ``fat jet'' with 
\beeq
\label{eq:fatjetcrit}
p_T(\hbox{fat jet})\geq 150\gev \quad \hbox{and}\quad |y(\hbox{fat jet})|<2\,,
\eeeq
defined via the inclusive Cambridge-Aachen
algorithm~\cite{Dokshitzer:1997in} 
with resolution parameter $R=1.2$. This particular choice of $R$
guarantees that we pick up the bulk of the $Z$ decay jets
since the boosted $Z\rig jj$ decay products' 
separation can be estimated to be $\sim 2 m_Z /p_T(Z)$.
Throughout, we invoke the algorithms and {\textsc C++} classes provided by the {\sc{FastJet}} 
framework~\cite{Cacciari:2005hq}.

The hadronic $Z$ reconstruction is performed applying the strategy of Ref.~\cite{Butterworth:2008iy}:
For the hardest jet in the event we undo the last stage of clustering, leaving two subjets, which we order with respect to
their invariant masses $m_{j_1} > m_{j_2}$. Provided a significant mass drop for a
not too asymmetric splitting, 
\beeq
\label{Eq:massdrops}
m_{j_1} < 0.67\, m_{j}\,,\;\;
\Delta R_{j_1j_2}^2 \text{min}(p_{T,j_1}^2, p_{T,j_2}^2) > 0.09\, m_j^2\,, 
\eeeq
where $\Delta R$ denotes the distance in the azimuthal angle--pseudorapidity plane,
we
consider the jet $j$ to be in the neighborhood of the resonance and terminate the declustering. Otherwise
we redefine $j$ to be equal to $j_1$ and continue the algorithm until the mass-drop condition is met. In case
this does not happen for the considered event, and we discard the event entirely.
If the mass-drop condition is met, we proceed with filtering of the fat jet~\cite{Butterworth:2008iy}; i.e. the constituents 
of the two subjets which survive the mass drop condition are recombined with higher 
resolution 
\bee
R_\text{filt} =\min\left(0.3\,, {\Delta R_{j_1j_2} \over 2}\right)\,,
\eee
and the three hardest filtered subjets are again required to
reproduce the $Z$ mass within $m_Z \pm 10~\text{GeV}$.

We subsequently reconstruct the Higgs mass from the excess in the $m_X^2 = (p_{Z_h} + p_{Z_\ell})^2$
distribution, i.e. we imagine a situation where the $X$ mass peak has already been established experimentally.
This allows us to avoid dealing with the sophisticated details of experimental strategies, which aim to single out the resonance 
peak from underlying event, pile-up and background distributions by typically including 
combinations of various statistical methods . 
A thorough discussion would be beyond the scope of this work. For a considered $X$ mass
of $400\gev$ we include events characterized by reconstructed invariant masses
\beeq
\left[p(\mu^+)+p(\mu^-)+ p(j_\alpha)+ p(j_\beta)\right]^2 = (400\pm 50 \gev )^2\,.
\eeeq
Further signal-over-background ($S/B$) improvements can be achieved by requiring
\be
\Delta R_{ZZ} < 3.2
\ee
and by trimming and pruning \cite{Ellis:2009su} of the hadronic $Z$ event candidates on the massless cell level 
of the event \cite{Soper:2010xk}, as described in detail in Ref.~\cite{Hackstein:2010wk}.

\section{Results and Discussion}
\label{sec:results}
In Figs.~\ref{fig:helhad}-\ref{fig:anglebkg} we show the angles of Eq.~\gl{eq:angledefinition} after 
various steps of the analysis have been carried out. We also give a comparison of the full hadron-level result
and Monte Carlo truth; i.e. we take into account the shower's particle information. These plots are the main results 
of this paper.
Comparing the two shower and hadronization approaches of {\sc{Pythia}} 6.4 and {\sc Herwig++}
for the process efficiencies in Table~{\ref{tab:cutscenarios}}, we find substantial discrepancies at intermediate steps of our analysis. 
After the entire analysis has been carried out this 
translates into a systematic uncertainty of $\sim 30\%$ of the total cross sections.
This is not a too large disagreement as both programs rely on distinct philosophies and approaches, which
typically result in sizable deviations when compared for identical Monte Carlo input.
The plots in Figs.~\ref{fig:helhad}-\ref{fig:anglebkg} show distributions obtained with {\sc{Herwig++}}.

We now turn to the discussion of the angular correlations.
It is immediately clear that the chosen selection criteria, Eqs.~\gl{eq:mupt}-\gl{eq:fatjetcrit}, do heavily affect the sensitive angular 
distributions of Eqs.~\gl{eq:angledefinition}. Retaining a signal-over-background ratio of approximately $0.5$, however, does not 
allow us to relax the $p_T$ cut on the fat jet. This cut turns out to be lethal to some of the angular distributions. 
Referring, e.g., to $\cos\theta^\star$, plotted in Fig.~\ref{fig:anglestar}, 
we find that our fat jet criteria, Eq.~\gl{eq:fatjetcrit}, force the distribution into reflecting extremely 
hard and central decay products. This removes essentially all discriminating features from the 
differential distribution $\d \sigma/\d\cos\theta^\star$, that show up for 
$|\hspace{-0.03cm}\cos\theta^\star| \gtrsim 0.5$ at the (inclusive) Monte Carlo event generation level. 
This is also reflected in the distinct 
acceptance level of the different $J^{\cal{CP}}$ samples, shown in Table~\ref{tab:cutscenarios}.
Note that, throughout, the fully hadronic distributions are in very good agreement with the Monte Carlo-truth
level.

Most of the sensitivity found in the observable $\tilde\Phi$ for the signal sample can be carried over to the
hadron level.
Yet, the angular pattern is known to be sensitive to the $X$'s mass scale, tending to decorrelate for
larger $X$ masses (see~e.g.~\cite{Gao:2010qx}).

As already pointed out in Sec.~ \ref{sec:details}, the ambiguity in $\cos\Phi$
smears out the angular correlations quite a lot in Fig.~\ref{fig:phi}. 
This comes not as too large
limitation of the angle's sensitivity for a ${\cal{CP}}$-odd scalar particle $X$. For $X=0^-$, the distribution peaks
at $\Phi=\pi/2$ and is also rather symmetrical with respect to $\pi/2$.
This leaves us after the subjet analysis with the helicity angles of Eq.~\gl{eq:costhetahel} and $\tilde\Phi$ as three 
sensitive angles out of five not taking into account the background distribution.

Crucial to obtaining angular correlations after all, is the analysis' capability to reconstruct both of the 
$Z$ rest frames (and from them the $X$ rest frame). This is already clear from the angles' definition 
in Eq.~\gl{eq:angledefinition}, and, again, this is not an experimental problem considering the 
purely leptonic channels.
For the angles $\Phi$ and $\theta^\star$
decorrelate (with the exception of $0^-$) due to the selection criteria, 
a bad rest frame reconstruction would not be visible in these
observables immediately. This is very different
if we turn to the helicity angles. Quite obviously, given a good hadronically decaying $Z$ rest frame reconstruction,
we can apply the {\emph{identical}} leptonic helicity angle as invoked for the measurement in 
$X\rig ZZ\rig 4\ell$; we have referred to this angle as $\theta_\ell$, previously. 
The only difference compared to the purely leptonic analysis is that we consult a partly hadronic system to construct
the reference system, in which the leptonic helicity angle $\theta_\ell$ is defined.

Indeed, the subjet analysis described in Sec.~\ref{sec:discr} 
is capable of giving a very good reconstruction of the hadronically decaying $Z$ boson
rest frame, while sufficiently reducing the backgrounds. 
This allows to carry over most of the central sensitivity of the angular distributions in Fig.~\ref{fig:hellep} to the fully simulated final state. 
However, the hadronically-defined helicity angle, displayed in Fig.~\ref{fig:helhad}, 
also suffers badly from the subjet analysis. Note that the bulk of the modifications of $\cos\theta_h$ 
do not arise from our restrictive selection criterion Eq.~\gl{eq:fatjetcrit}, but from symmetry requirements 
among the subjets in the mass-drop procedure. Thus, the subjets which provide a significant mass drop 
with respect to Eq.~\gl{Eq:massdrops} are biased towards $\theta_h \simeq 90 ^{\circ}$.

A remaining key question that needs to be addressed is whether the potentially sensitive angles $\theta_\ell, \Phi$, and $\tilde \Phi$ 
exhibit visible spin- and ${\cal{CP}}$-dependent deviations when the background distribution is taken into account. 
We show these angles including the backgrounds in Fig.~\ref{fig:anglebkg}.
The backgrounds' $\tilde \Phi$ distribution largely mimics the $1^+$ shape under the subjet analysis' conditions, so we cannot 
claim sensitivity unless the backgrounds distribution is very well known.
This also accounts for the $\Phi$ distribution in a milder form. 
While here the background is flat to good approximation, $S/B$ (see Table~\ref{tab:cutscenarios}) limits the sensitivity to the 
shape deviations, which are ameliorated due to the different signal efficiencies. However, the distribution remains sensitive to $X=0^-$ shape.
$pp\rig X \rig \mu^+\mu^- jj$ remains
sensitive to the ${\cal{CP}}$ quantum number of a scalar particle $X$ in the $\cos\theta_\ell$ distribution, which
is opposite in shape compared to the background distribution.
 
We have only considered an $X$ mass $m_X=400\gev$, a choice which is quite 
close to the lower limit of the mass range, where the boosted analysis is applicable.
Some remarks concerning our analysis
for different $X$ masses and widths are due. The boost requirements and the centrally required
selection cuts do affect the angular distributions in a $X$ mass-independent manner. The remaining
angles are then qualitatively determined by the goodness of reconstruction,
which becomes increasingly better for heavier $X$ masses, keeping the width fixed.
 
In case of the SM Higgs boson, the width is proportional to $m_H^3$ due to
the enhanced branching of the Higgs to longitudinally-polarized $Z$s. With the resonance 
becoming width-dominated, our mass reconstruction still remains sufficiently effective; $S/B$, 
however, increasingly worsens.
For these mass ranges, the analysis is sensitive to the experimental methods that recover the resonance
excess. Additionally, from a theoretical perspective, there are various models known in the literature where a heavy
resonance becomes utterly narrow or exceedingly broad (see, e.g. \cite{Birkedal:2004au,Bagger:1995mk} for
discussions of the resulting phenomenology). 
The former yields, depending on the (non-SM) production cross section, a better mass reconstruction,
while the latter case is again strongly limited by $S/B$; cf. Fig.~\ref{fig:anglebkg}.
For any of these EWSB realizations, our methods should be modified accordingly, taking into account all realistic 
experimental algorithms, techniques, and uncertainties as well as all model-dependent parameters.

\section{Summary and Conclusions}
\label{sec:conc}
The discovery of a singly produced new state at the LHC, and methods
to determine its additional quantum numbers, remains an active field of 
particle physics phenomenology research.
The availability of tools to simulate events at a crucial level
of realism and precision has put the phenomenology community into a position
that allows a more transparent view onto the complicated particle dynamics at 
high-energy hadron colliders than ever. 

In this paper we have
explored the performance of new jet techniques when applied to the analysis of spin- and ${\cal{CP}}$-sensitive 
distributions of a newly discovered resonance, which resembles the SM in 
the overall rate in $pp\rig \mu^+\mu^-jj$.
We have performed a detailed investigation of the angular correlations
and have worked out the approach-specific limitations, resulting from the boosted and central
kinematical configurations. It is self-evident that a QCD-dominated final state cannot compete
with a leptonic final state in terms of signal purity, higher order and shower uncertainties, {\emph{per se}}.
These uncertainties are inherent to any current discussion related to jet physics. Nonetheless, we have shown
that potential ``no-go theorems'' following from huge underlying event and QCD background rates 
for $pp\rig X \rig ZZ \rig \ell^+\ell^- j j$ can be sufficiently ameliorated to yield an overall 
sensitivity to the ${\cal{CP}}$ property of a singly produced scalar resonance. 
Straightforwardly applying the described analysis strategy to vectorial and tensorial resonances 
does not yield reliable shape deviations when the backgrounds' distribution is taken into account.
Given that the cross section of the semihadronic decay channel is approximately 10 times
larger compared to $X\rig 4\ell$, the performed subjet analysis qualifies to at least supplement
measurements of the purely leptonic decay channels.

A question we have not addressed in this paper is the potential application of the presented strategy 
to signatures, which do not resemble the SM at all. Electroweak symmetry breaking 
by strong interactions is likely to yield a large rate of longitudinally polarized electroweak bosons 
due to modified  $XZZ$ and $X\bar q q$ couplings \cite{Chanowitz:1985hj,silh}. 
Measuring the fraction of longitudinal polarizations, which can be inferred from the $Z$'s decay
products' angular correlation as proposed recently in Ref.~\cite{Han:2009em}, should benefit from 
the methods we have investigated in this paper. This is in particular true for new composite operators, such as a 
modification of the Higgs kinetic term \cite{silh}, inducing
asymmetric angular decay distributions of the leptons. In addition, our analysis is also applicable 
to the investigation of isovectorial resonances (see, e.g., Ref.~\cite{Bagger:1995mk}) 
in $pp \rig WZ$, with the $W$ decaying to hadrons, and $Z\rig \mu^+\mu^-$. 
We leave a more thorough investigation of these directions to future work.
\vskip0.1\baselineskip

{\bf{Acknowledgments}} ---
We thank Tilman Plehn for many discussions and valuable comments on the manuscript.
This work was supported in part by the US Department of Energy under Contract No. 
DE-FG02-96ER40969. C.H. is supported by the Graduiertenkolleg 
``High Energy Particle and Particle Astrophysics''. 
Parts of the numerical calculations presented
in this paper have been performed using the Heidelberg/Mannheim and Karlsruhe high performance clusters of the
bwGRiD ({\tt http://www.bw-grid.de}), member of the German D-Grid initiative, funded by the Ministry for Education and 
Research (Bundesministerium f\"ur Bildung und Forschung) and the Ministry for 
Science, Research and Arts Baden-Wuerttemberg (Ministerium f\"ur Wissenschaft, 
Forschung und Kunst Baden-W\"urttemberg). C.H. thanks M.~Soysal for the support using the Karlsruhe cluster.

\end{document}